\documentclass[prb,aps,nofootinbib, twocolumn, amssymb,superscriptaddress,10pt]{revtex4-2}
\usepackage{amsmath}
\usepackage[colorlinks,citecolor=blue,linkcolor=blue,urlcolor=blue]{hyperref}
\usepackage{graphicx}
\usepackage{booktabs}
\usepackage{xcolor}
\usepackage{stix}

\begin{document}

\title{Machine Learning for Electron-phonon Interactions From Finite Difference}


\author{Zun Wang}
\affiliation{State Key Laboratory of Low Dimensional Quantum Physics and Department of Physics, Tsinghua University, Beijing 100084, China}
\author{Wenhui Duan}
\affiliation{State Key Laboratory of Low Dimensional Quantum Physics and Department of Physics, Tsinghua University, Beijing 100084, China}
\affiliation{Institute for Advanced Study, Tsinghua University, Beijing 100084, China}
\affiliation{Frontier Science Center for Quantum Information, Beijing 100084, China}
\author{Zuzhang Lin}
\email{linzz@hku.hk}
\affiliation{New Cornerstone Science Lab, Department of Physics, The University of Hong Kong, Hong Kong, China}
\affiliation{HK Institute of Quantum Science \& Technology, The University of Hong Kong, Hong Kong, China}
\affiliation{State Key Laboratory of Optical Quantum Materials, The University of Hong Kong, Hong Kong, China}

\date{\today}%

\begin{abstract}
First-principles investigations of electron-phonon interactions (EPIs) play a crucial role in understanding a wide range of phenomena in physics and materials science.
Among various approaches, the finite difference method offers a direct route to capture higher-order EPIs and is compatible with diverse electronic structure solvers. However, its considerable computational cost limits its broader application.
To overcome this bottleneck, we present a machine learning electron-phonon interaction (MLEPI) pipeline that predicts force constants and electronic Hamiltonians for modeling EPIs from finite difference calculations, improving efficiency by orders of magnitude without compromising accuracy.
The performance of MLEPI is validated by studying the temperature dependence of the electronic band properties in bilayer graphene, where both first- and second order EPIs are treated on an equal footing.
Using a heterogeneous edge network, the pipeline integrates both interlayer and intralayer interactions, making it particularly suitable for studying multilayer materials.
With its inherent adaptability and ease of transfer to other applications, our methodology provides a robust tool with a very favorable accuracy/efficiency balance for investigating EPIs in large-scale material systems.

\end{abstract}

\maketitle

\section{Introduction}
Electron-phonon interactions (EPIs) are fundamental to a wide range of physical phenomena, such as charge and heat transport, phonon-mediated superconductivity, phonon-assisted optical absorption, and so on. First-principles methods enable quantitative calculations of EPI effects \cite{giustino2017electron,monserrat2018electron}. Primarily, two approaches are employed for this purpose: the linear response method, which requires extensive calculations of EPI matrix elements in the Brillouin zone, and the finite difference method, which involves simulating explicit atomic displacements in large supercells commensurate with all phonon wave vectors. 
Both approaches face substantial computational challenges, particularly when applied to large-scale or structurally complex systems.
Therefore, it is essential to improve both the computational efficiency and precision of EPI calculation methods, thereby enabling their application to increasingly complex material systems.

Compared to the linear response approach, the finite difference approach offers several distinct advantages. 
One significant advantage is its universal compatibility with electronic structure solvers and thus it is feasible to include strong electron correlations when studying EPIs \cite{antonius_many-body_2014, monserrat_correlation_2016}. 
Another advantage is its inherent ability to seamlessly integrate EPIs beyond the lowest order.  
The significance of high-order EPIs has been underscored \cite{lee2020ab,alldredge1967role,thorbergsson1986mobility,sher1967resonant,woods1998nonlinear,wang1989low,story2014cumulant,nery2018quasiparticles,zhou2019predicting}. 
An example is the equal importance of the first- and second-order EPI terms in addressing temperature-dependent electronic structures \cite{allen_theory_1976}.
Moreover, the finite difference approach enables exploring systems beyond the harmonic approximation \cite{monserrat2018electron}. 
These advantages position the finite difference method as a preferred choice for exploring EPIs \cite{lin_phonon-limited_2022}, provided that its substantial computational demands can be effectively addressed.
Although considerable research efforts have been devoted to this goal \cite{lloyd-williams_lattice_2015, chen_nonuniform_2022,patrick_quantum_2013,monserrat_vibrational_2016,zacharias_one-shot_2016,monserrat_correlation_2016-1}, the acceleration of the finite difference methods has thus far only been achieved to a limited extent.

Machine Learning (ML) offers an unparalleled opportunity to address the computational challenges associated with the finite difference method. Recently, it has experienced substantial progression~\cite{jordan2015machine} and has been harnessed in the realm of AI for Science~\cite{wang2023scientific}, resulting in a notable acceleration in computations related to Density Functional Theory (DFT). 
At the atomic level, a prominent line of research involves machine learning force fields (MLFFs) ~\cite{behler2007generalized, schutt2017schnet, schutt2018schnet, thomas2018tensor, coors2018spherenet, anderson2019cormorant, gasteiger2019directional, gasteiger2020fast, fuchs2020se, schutt2021equivariant, gasteiger2021gemnet, tholke2021equivariant, batzner2022, musaelian2023learning, liao2022equiformer, batatia2022mace, wang2022comenet, wang2023efficiently, liao2023equiformerv2, wang2024enhancing, li2024longshortrange, wu2024se3set}, which aim to predict the forces and potential energy surfaces of molecules from atomic interactions. At the electronic level, pioneering studies~\cite{schutt2019unifying, unke2021se, li2022deep, gong2023general, yu2023efficient, zhangself, wang2024infusing,li_neural-network_2024} have employed graph neural networks (GNNs) to model the electronic Hamiltonian. 
More recently, initial efforts have been made to incorporate ML into the modeling of EPIs within the framework of linear response theory~\cite{li2024deep, zhong2024accelerating}. 
However, the application of machine learning to model EPIs using the finite difference approach, which could naturally accommodate higher-order interactions, strong electron correlations, and anharmonic effects, remains largely unexplored.


\begin{figure*}[htbp]
\includegraphics[width=1.0\textwidth]{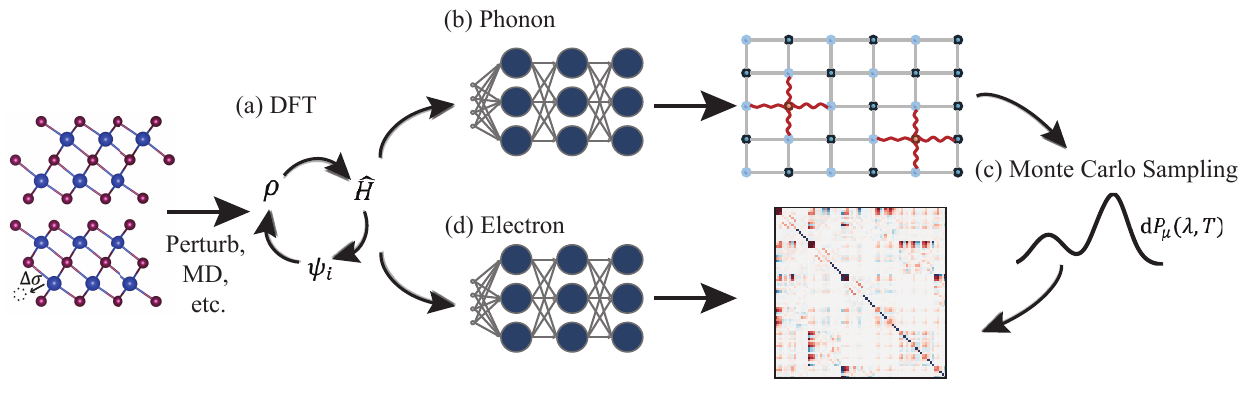}
  \centering
  \caption{MLEPI pipeline for ML-based EPI calculations in materials. For an arbitrary material, (a) DFT is employed to generate phonon training sets that contain structural information, energies and atomic forces, as well as electronic training sets that include structural details and Hamiltonians.  These datasets are used to train (b) a phonon neural network and (d) an electronic neural network, respectively. The trained phonon neural network predicts the interatomic forces for perturbed structures, facilitating the calculation of phonon-eigenmode determined probability distributions $dP_{\mu}(\lambda, T)$ required for (c) MC sampling. The structures sampled via MC are then input into the trained electronic neural network to infer corresponding Hamiltonians, which are instrumental for subsequent calculations of EPI-related physical quantities. $\Delta \sigma$ schematically represents the perturbation of the atomic structure around the equilibrium configuration.}
  \label{fig:1}
\end{figure*}

In this work, we introduce a comprehensive GNN-based framework, termed the machine learning electron-phonon interaction (MLEPI) pipeline, which integrates ML algorithms of predicting force constants with those proficient in predicting DFT Hamiltonians. This approach enables efficient and accurate investigations of EPIs in large-scale material systems.
Especially we employ heterogeneous graph neural networks, termed the heterogeneous edge network (HedgeNet), to explicitly capture both interlayer and intralayer interactions in multilayer materials. This capability is particularly important, as the interlayer interaction plays a pivotal role in the emergent properties of multilayer systems — an area of intense interest in condensed matter physics and materials science.
Importantly, HedgeNet leverages the finite difference method to model EPIs, naturally incorporating higher-order EPI effects that are typically beyond the reach of existing ML methods focused primarily on the lowest-order EPIs. We have conducted experiments on bilayer graphene to evaluate the performance of HedgeNet and demonstrate its robustness. 
Our work thus establishes a scalable and accurate ML-based GNN framework for investigating EPIs across a broad spectrum of materials.

The remainder of this article is organized as follows. Section II develops the MLEPI pipeline framework, describes the implementation of HedgeNet, and evaluates its capabilities. Section III applies the MLEPI pipeline in experiments on bilayer graphene, conducts ablation studies, and analyzes HedgeNet’s efficiency, followed by a discussion of related work and an assessment of computational complexity. Section IV provides the conclusions.


\section{MLEPI pipeline}
\subsection{Conceptual Framework of the MLEPI pipeline}
The finite difference method offers a powerful first-principles approach for calculating EPI effects and predicting various related phenomena. It computes the electronic response to the explicit displacements of atomic nuclei. In the adiabatic approximation, the expectation value of an operator at temperature $T$ is $A(T) = \frac{1}{\mathcal{Z}} \sum_k\left\langle\chi_k(\mathbf{R})\right| \mathcal{A}(\mathbf{R})\left|\chi_k(\mathbf{R})\right\rangle \mathrm{e}^{-E_k / k_{\mathrm{B}} T}$ with $\mathcal{A}(\mathbf{R})=\langle\psi(\mathbf{r} ; \mathbf{R})| \mathcal{A}(\mathbf{r}, \mathbf{R})|\psi(\mathbf{r} ; \mathbf{R})\rangle$, where $\left|\chi_k(\mathbf{R})\right\rangle$ is the nuclear state, $|\psi(\mathbf{r} ; \mathbf{R})\rangle$ is the electronic state, $E_k$ is the total energy of the electron-nucleus system in state $k$, $\mathcal{Z}=\sum_k \mathrm{e}^{-E_k / k_{\mathrm{B}} T}$ is the partition function, $k_B$ is Boltzmann's constant, $\mathbf{r}=\{\mathbf{r}_i\}$ denotes collective electron coordinate, and $\mathbf{R}=\{\mathbf{R}_{\alpha}\}$  specifies the coordinates of the nuclei, which are treated as parameters. 
Evaluating this expectation value requires electronic structure calculations across all possible $\mathbf{R}$ configurations, which is computationally intractable. In crystalline solids, however, atomic nuclei typically oscillate with small amplitudes around their equilibrium positions, and thus the harmonic approximation generally holds. Under the harmonic approximation, $|\chi\rangle$ is replaced by its harmonic counterpart and the $A(T)$ can be evaluated using  Monte Carlo (MC) sampling scheme \cite{monserrat2018electron}
\begin{equation}
    \langle A(T)\rangle = \frac{1}{m}\sum_{j=1}^mA(y_T^{\text{MC}, j}), \label{eq:obs}
\end{equation}
where $y_T^{\text{MC}, j} = y_{\text{eq}} + \Delta \sigma^{\text{MC}, j}$, $y_{\text{eq}}$ is the equilibrium atomic position, and $\Delta \sigma^{\text{MC}, j} = \sqrt{\frac{1}{M_{\lambda}}}\sum_{\mu}^{3(N-1)}\epsilon_{\lambda, \mu}\mathcal{N}$ is the displacement, whose amplitude is obtained from the normally distributed random variable $\mathcal{N}$ with a probability distribution~\cite{karsai2018electron}:
\begin{equation}
    dP_{\mu}(\lambda, T) = \frac{1}{\sqrt{2\pi\langle v_{\mu\lambda}^2\rangle}}\exp\left(-\frac{\lambda^2}{2\langle v_{\mu\lambda}^2\rangle}\right)d\lambda. \label{eq:prob}
\end{equation}
Here, $N$ is the atom number, $\epsilon_{\lambda \mu}$ denotes the unit vector of eigenmode $\mu$ on atom $\lambda$ ($\lambda$ includes both the Cartesian coordinates and the atom index), $M_{\lambda}$ and $\omega_\mu$ respectively represent the atom mass and phonon eigenfrequency, and $\langle v_{\mu\lambda}^2\rangle = \frac{\hbar}{2M_{\lambda}\omega_{\mu}}\coth{\frac{\hbar\omega_{\mu}}{2k_BT}}$ is the mean squared displacement of the harmonic oscillator.

In the conventional approach, computing $\langle A(T)\rangle$ in Equation \ref{eq:obs} begins with the determination of the phonon eigenmodes, followed by the evaluation of the expectation value across all distributed MC configurations. Electronic structure calculations are required at every step of this workflow. This process is computationally demanding, as it necessitates numerous calculations of electronic structures for supercells that are commensurate with the phonon wave vectors. And the supercell can be exceptionally large, particularly when long-wavelength phonons are involved.  

Our proposed MLEPI pipeline aims to address this computationally demanding issue. As shown in Fig.~\ref{fig:1}, the entire workflow can be divided into a phonon part and an electron part, connected by MC sampling. For the phonon part, we use an MLFF for atom-level material characterization. By incorporating equivariance (refer to Section A in Supplementary Material (SM) for more information on related equivariant concepts\textbf{}) into ML methods~\cite{behler2007generalized, schutt2017schnet, schutt2018schnet, thomas2018tensor, coors2018spherenet, anderson2019cormorant, gasteiger2019directional, gasteiger2020fast, fuchs2020se, schutt2021equivariant, gasteiger2021gemnet, tholke2021equivariant, batzner2022, musaelian2023learning, liao2022equiformer, batatia2022mace, wang2022comenet, wang2023efficiently, liao2023equiformerv2, wang2024enhancing, li2024longshortrange, wu2024se3set}, MLFF can accurately predict interatomic forces, energies, and other properties of molecules and materials at a fraction of the computational cost of \textit{ab initio} methods, which are typically computationally expensive and time-consuming. 
For the electron part, DFT is one of the most powerful traditional methods for constructing effective Hamiltonians to predict material properties. It simplifies the complex many-body problem by transforming it into an auxiliary problem, $\hat{H}_{\rm{DFT}}\left|\psi\right\rangle = \mathcal{E}\left|\psi\right\rangle$.
In this equation, $\hat{H}_{\rm{DFT}}$ is the DFT Hamiltonian operator, and $\mathcal{E}$ and $\left|\psi\right\rangle$ are the Kohn–Sham eigenvalue and eigenstate, respectively. While DFT typically exhibits excellent predictive capabilities in small-scale systems, it becomes computationally demanding and its effectiveness significantly decreases when applied to large-scale systems. Alternatively, ML methods~\cite{schutt2019unifying, unke2021se, li2022deep, gong2023general, yu2023efficient, zhangself, wang2024infusing, gong2024generalizing, tang2024deep} offer an efficient approach, capable of predicting the Hamiltonian with high accuracy and significantly lower computational cost. 

The MLEPI pipeline, illustrated in Fig.~\ref{fig:1}, is a general framework applicable to a wide range of material systems.
Firstly, training data (including energies, forces, and Hamiltonian matrix elements, etc.) are generated using DFT, in alignment with the requirements of the MLFF and Hamiltonian predictor.
These labeled datasets are then used to independently train the phonon network and the electron network. During inference, the phonon network predicts interatomic forces for a given structure, enabling the calculation of phonon eigenmodes. Based on these, the probability distribution $dP_{\mu}(\lambda, T)$ is computed according to Equation~\ref{eq:prob}. Multiple structures are then sampled from this distribution and fed to the electron network, which predicts the corresponding Hamiltonian. As a result, physical observables modulated by EPIs can be computed with high efficiency and accuracy.

\begin{figure}[htbp]
\includegraphics[width=0.5\textwidth]{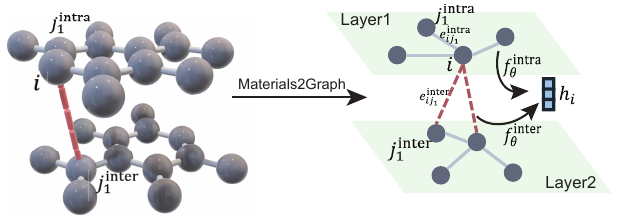}
  \centering
  \caption{A schematic illustrates how a multi-layered material system can be constructed as a heterogeneous graph $\mathcal{G} = (\mathcal{V}, \mathcal{R}, \mathcal{E})$. Without loss of generality, we consider a bilayer material and arbitrarily select a central atom $i$ from Layer 1. Its neighboring atoms $j \in \mathcal{N}_i$ are identified based on a specified cutoff radius. In the graph, the superscript of $j$, either intra or inter, indicates whether it is in the same layer or a different layer compared to atom $i$. The resulting heterogeneous graph encompasses two types of relations: $\mathcal{R} = \{\text{rel}_{\text{intra}}, \text{rel}_{\text{inter}}\}$. The connecting edge  $e_{ij}^{\text{intra}}$ (represented by a solid gray line) signifies an intra-layer relationship $\text{rel}_{\text{intra}}$, while the connecting edge $e_{ij}^{\text{inter}}$ (shown as an orange dashed line) signifies an inter-layer relationship $\text{rel}_{\text{inter}}$. For the central atom $i$, edges corresponding to the intra-layer relationship $\text{rel}_{\text{intra}}$ are input into the neural network module $f_{\theta}^{\text{intra}}$ (where $\theta$ denotes learnable parameters), while edges corresponding to the inter-layer relationship $\text{rel}_{\text{inter}}$ are input into $f_{\theta}^{\text{inter}}$. This process yields the node embedding $h_i$ for the central atom $i$.}
  \label{fig:2}
\end{figure}

\subsection{HedgeNet}
Van der Waals (vdW) materials \cite{bhimanapati2015recent}, composed of atomically thin layers held together by weak vdW interactions, offer a versatile platform for investigating emergent quantum phenomena, with electronic, optical, and magnetic properties that can be precisely tuned. 
In these materials, intralayer and interlayer interactions exhibit substantial differences. While intralayer interactions, including covalent and ionic bonds, are primarily dominated by strong short-range effects and govern the intrinsic electronic properties, the typically long-range interlayer vdW interactions allow for adjustable stacking arrangements—such as twisting, sliding, and strain—enabling precise modulation of quantum phases. Accordingly, independently treating intralayer and interlayer interactions enables more efficient and accurate calculations of EPIs in these materials.

\begin{figure*}[htbp]
\includegraphics[width=1.0\textwidth]{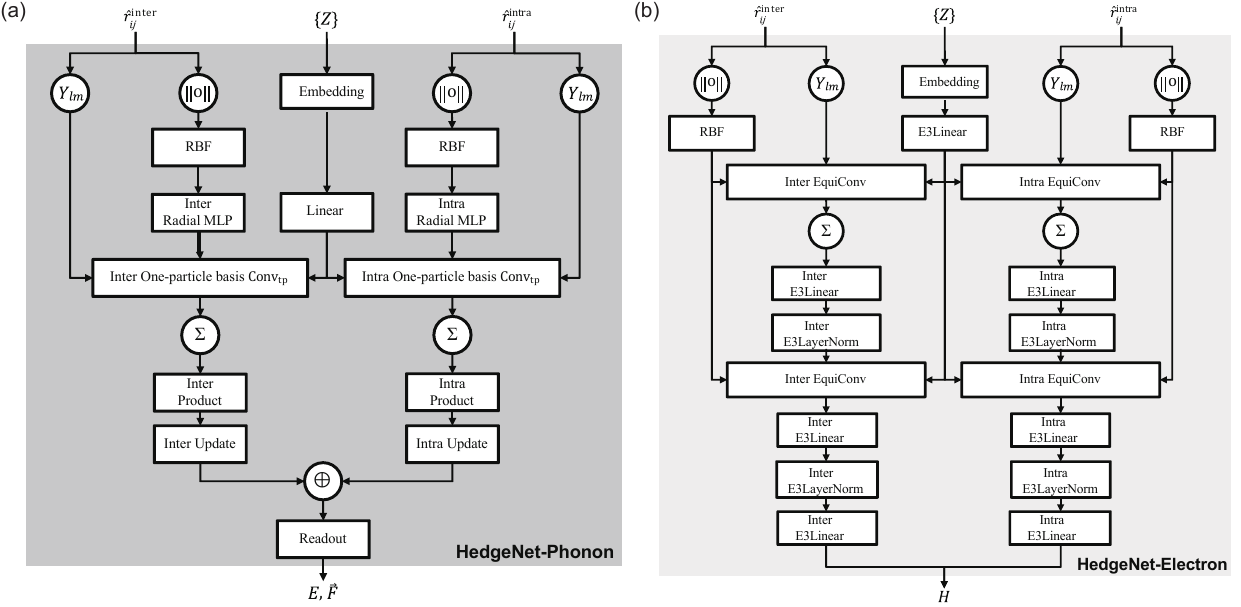}
  \centering
  \caption{The detailed architecture of (a) HedgeNet-Phonon and (b) HedgeNet-Electron. In this work, MACE~\cite{batatia2022mace} and DeepH-E3~\cite{gong2023general} are utilized as the backbones for the phonon network (HedgeNet-Phonon) and electronic network (HedgeNet-Electron), respectively. The architectural details of the vanilla MACE and DeepH-E3 models can be found in Sections B and C in SM, respectively.}
  \label{fig:3}
\end{figure*}

In the MLEPI pipeline, we propose HedgeNet to introduce heterogeneous graph neural networks (HGNNs) for addressing intralayer and interlayer interactions independently.
The implementation of HGNNs, which are grounded in GNNs, is applied in both the phonon and electron components of the pipeline. The core concept of GNNs can be summarized as a message-passing neural network (MPNN)~\cite{gilmer2017neural}, in which each node iteratively updates its embedding by aggregating messages from its neighboring nodes. 
In a graph $\mathcal{G} = (\mathcal{V}, \mathcal{E})$, MPNN can be summarized as
\begin{equation}
    h_v^{(l+1)} = U^{(l)}\left(h_v^{(l)}, \sum_{w \in \mathcal{N}_v} M^{(l)}(h_v^{(l)}, h_w^{(l)}, e_{vw}^{(l)})\right),
\end{equation}
where $\mathcal{V}$ represents a set of nodes, $\mathcal{E}$ is a set of edges, $h_v$ and $e_{vw}$ respectively denote node and edge features, $M^{(l)}$ and $U^{(l)}$ are respectively the message function and update function of $l$-th layer, $\mathcal{N}_{v}$ denotes the set of neighbors of node $v$, and $u$ and $v$ are indices of nodes. 
In a heterogeneous graph $\mathcal{G} = (\mathcal{V}, \mathcal{R}, \mathcal{E})$, the generalized forward process is updated to
\begin{equation}
    h_i^{(l+1)} = \sigma \left( \sum_{\text{rel} \in \mathcal{R}} \sum_{j \in \mathcal{N}_i^{\text{rel}}} W_{\text{rel}}^{(l)} h_j^{(l)} + W_0^{(l)} h_i^{(l)} \right),
\end{equation}
where $\mathcal{R}$ denotes the set of relations, $\mathcal{N}_i^{\text{rel}}$ and $\sigma$ denote the set of neighbors of vertex $i$ for relation $\text{rel}$ and activation functions, respectively. $W_{\text{rel}}^{(l)}$ and $W_0^{(l)}$ are the learnable parameters in the neural networks.

The proposed HedgeNet is shown in Fig.~\ref{fig:2}. We construct a multilayered material as a heterogeneous graph based on its interlayer and intralayer interactions. For a central atom that is connected to both intralayer atoms and interlayer atoms, it will separately obtain node features from the intra block and the inter block. The node representations from these two blocks will be pooled by summing them to obtain the final representation of the central atom. In this work, we utilized MACE~\cite{batatia2022mace} and DeepH-E3~\cite{gong2023general} as the backbones for the phonon network (HedgeNet-Phonon) and electron network (HedgeNet-Electron), respectively. The architectural details of the original MACE and DeepH-E3 models can be found in Sections B and C in SM, respectively. We modified them into HGNNs to better simulate multilayer materials (see Fig.~\ref{fig:3}). Specifically, for a multilayer material, there are two dual sub-networks that respectively model the interlayer and intralayer atomic interactions. At the end of the network, the embeddings learned for interlayer and intralayer interactions are combined. A complexity analysis of this design will be provided in the discussion section.

\subsection{Capability of the MLEPI pipeline}
The developed MLEPI pipeline (Fig.~\ref{fig:1}) is universal for accelerating the calculations of EPI effects on diverse physical properties—such as band structure, band geometry, and responses to external fields—while circumventing DFT self-consistent calculations. It is applicable to a broad spectrum of materials, demonstrating particularly good performance in multilayer materials.
We validate our approach using bilayer graphene, focusing on EPI‑renormalized electronic properties that require second‑order EPIs, thereby underscoring the advantages of the MLEPI pipeline that is rooted in a finite difference approach. 
Furthermore, we present ablation studies to illustrate the performance of HedgeNet and engage in discussions about the high efficiency of our methodology, related works, and computational complexity.


\begin{figure*}[htbp]
\includegraphics[width=1.0\textwidth]{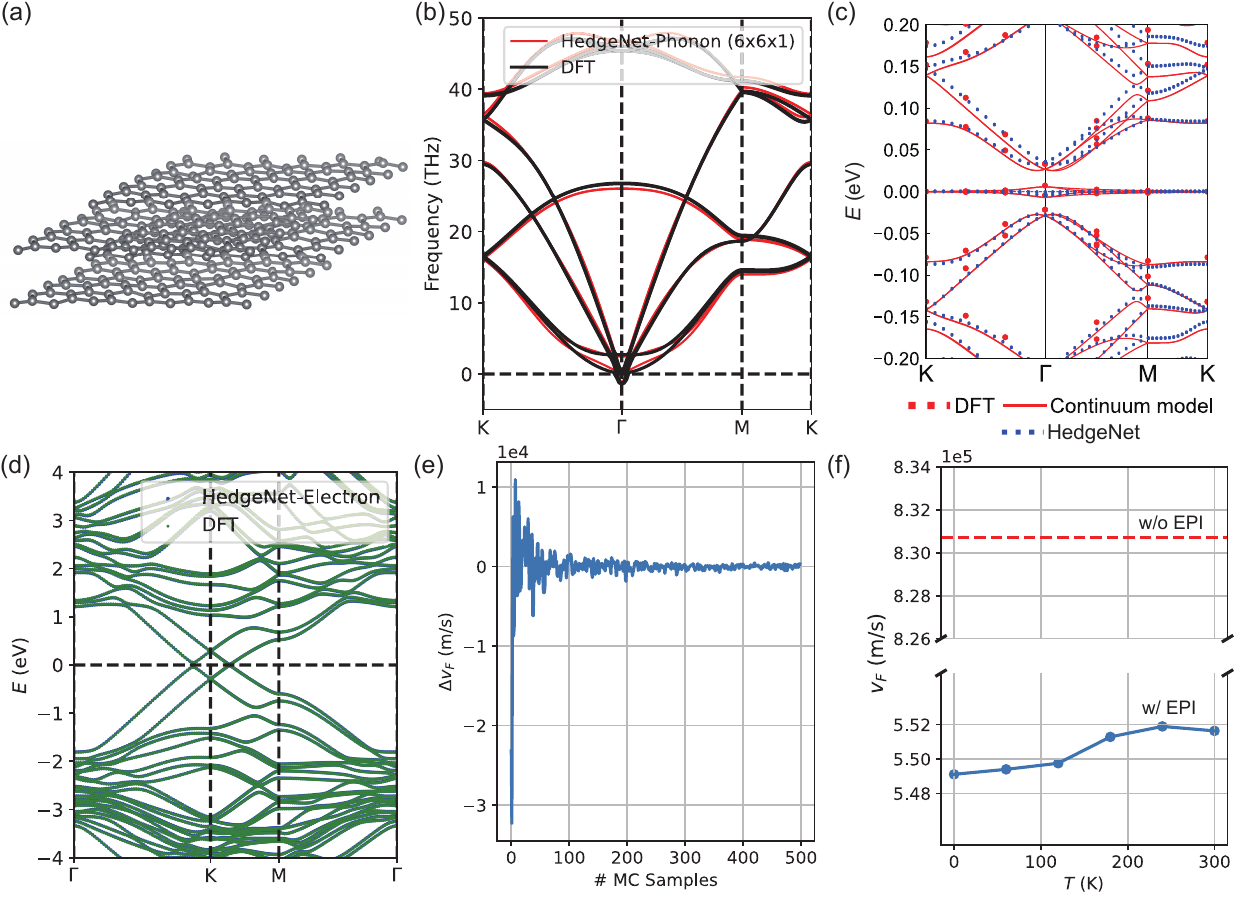}
  \centering
  \caption{(a) Schematic structure of bilayer graphene. (b) Comparison of the phonon spectrum of a 6$\times$6$\times$1 bilayer graphene supercell predicted by HedgeNet-Phonon model, trained on a 4$\times$4$\times$1 bilayer graphene dataset, with DFT results. (c) Comparison of the band structure inference results of twisted bilayer graphene (magic angle $\theta=$1.08$^\circ$, 11,164 atoms) from HedgeNet-Electron, trained on the bilayer graphene dataset provided by DeepH~\cite{li2022deep} and DeepH-E3~\cite{gong2023general}, with DFT and continuum model results (The DFT and continuum model results are extracted from Ref.~\cite{lucignano2019crucial, li2022deep}, and~\cite{gong2023general}). (d) Comparison of the band structure results of a randomly selected MC sampling configuration from HedgeNet-Electron and DFT. 
  (e) Iteration error of the Fermi velocity of the bilayer graphene modulated by EPI effects at 0 K, calculated according to the Equation~\ref{eq:obs}, as a function of the number of sampling points. The converge trendency is similiar for other temperatures. (f) Variation of the Fermi velocity of the bilayer graphene modulated by EPI effects with temperature, calculated with 500 sampling points. The dashed red line indicates Fermi velocity calculated without accounting for EPIs.
  }
  \label{fig:4}
\end{figure*}

\section{Application and Evaluation of the MLEPI Pipeline}
\subsection{Experiments of bilayer graphene}
Bilayer graphene was initially selected as the experimental system. Recognized as a groundbreaking one-atom-thick material, graphene exhibits remarkable characteristics in both its single- and multilayer forms, thereby establishing it as a transformative component across various disciplines, particularly in physics and materials science. We selected Fermi velocity $v_F$ as the physical quantity associated with the phonon-renormalized band, which is a crucial parameter in a variety of physical phenomena. 

To generate the training dataset for MLFF, 2,000 bilayer graphene structures with 4 by 4 supercells are produced, with minor random distortions from the equilibrium configuration.
The self-consistent electronic calculations of these structures are performed through DFT implemented in the VASP \cite{kresse1993ab,kresse1996computational}, employing projector-augmented wave potentials \cite{kresse1999ultrasoft} for the ion-electron interaction and generalized gradient approximation of Perdew-Burke-Ernzerhof \cite{perdew1996generalized} (PBE) for the exchange-correlation functional. SOC is not included and a 5$\times$5$\times$1 $k$-point mesh is used for all calculations. To include the van der Waals interaction, the dispersion-corrected DFT method (DFT-D2) \cite{grimme2006semiempirical} is employed. 
The periodic slabs are separated by more than 10 \AA. The self-consistent electronic calculations are stopped when the energy convergence reaches $10^{-6}$ eV. 
The plane-wave energy cutoff is 400 eV.
The DFT electronic structures are calculated via OpenMX software package~\cite{ozaki2003variationally, ozaki2004numerical}, using norm-conserving pseudopotentials and the PBE exchange-correlation energy functional with C6.0-s2p2. 
The adopted k-point mesh is $6 \times 6 \times 1$.
In all calculations, the cutoff energy is 200 Rydberg, the energy convergence criterion is $10^{-8}$ Hartree and SOC is excluded. 
The implementation details of HedgeNet, as well as the parameter settings for HedgeNet used in all experiments in this paper, can be found in Section D of SM.

We started by perturbing a 4$\times$4$\times$1 supercell of the equilibrium structure of the bilayer graphene (Fig.~\ref{fig:4}), including perturbations of the atomic coordinates within each layer and relative translations between different layers, ultimately obtaining 2,000 configurations. We then performed SCF calculations on these 2,000 configurations using Vienna ab initio simulation package (VASP) \cite{kresse1993ab,kresse1996computational} to obtain the corresponding energies and atomic forces. These data were randomly divided into training, validation, and test sets in the ratios of 0.81, 0.09, and 0.1, respectively, and were used to train and test the performance of the HedgeNet-Phonon. 
To train the HedgeNet-Electron model, we directly used Hamiltonian data for bilayer graphene provided by DeepH~\cite{li2022deep} and DeepH-E3~\cite{gong2023general}, adhering to their data splitting ratio, with the training, validation, and test sets divided in a 6:2:2 ratio.

The generalization capability of the trained HedgeNet-Phonon was tested by comparing the phonon spectrum of a 6$\times$6$\times$1 bilayer graphene structure with the results obtained from DFT calculations, as shown in Fig.~\ref{fig:4}(b). It is evident that HedgeNet-Phonon results align well with the DFT results.
We then tested the generalizability of the trained HedgeNet-Electron on twisted bilayer graphene (magic angle $\theta=$1.08$^\circ$, 11,164 atoms)~\cite{lucignano2019crucial}.
Figure~\ref{fig:4}(c) illustrates that the results from HedgeNet-Electron are in good agreement with those from DFT and the continuum model.

To study the EPI effects on Fermi velocity in bilayer graphene, the MC sampling was performed on a 6$\times$6$\times$1 bilayer graphene supercell based on the probability distribution corresponding to the phonon eigenmodes provided by HedgeNet-Phonon. This sampling was conducted at temperatures of 0 K, 60 K, 120 K, 180 K, 240 K, and 300 K, with 500 configurations generated for each temperature. These configurations were then used for the inference of HedgeNet-Electron. To further ensure the accuracy of HedgeNet-Electron, we randomly selected one of the MC-sampled configurations and performed DFT 
calculations on its band structure for comparison with the results from HedgeNet-Electron.
As shown in Fig.~\ref{fig:4}(d), the results from both methods are in perfect agreement. 

We then compute the EPI-renormalized Fermi velocity of bilayer graphene without doping as a function of temperature using Equation~\ref{eq:obs} (more details in Section E in SM). We first examine convergence with respect to the number of MC samples. As shown in Fig.~\ref{fig:4}(e), the results converge after a few hundred samples. 
At zero temperature, EPIs reduce the Fermi velocity of bilayer graphene by approximately 34\% [Fig.~\ref{fig:4}(f)], underscoring their significant influence on its electronic properties. While the linear-response approach predicts a considerably smaller reduction of 20\% \cite{attaccalite_doped_2010,sohier_phonon-limited_2014,example}, our ML based finite difference method reveals that higher-order EPI processes play crucial roles in the renormalization of electronic properties. Regarding temperature dependence, the EPI-induced modification of the Fermi velocity remains almost unchanged over the range of 0–300K.

\subsection{Ablation studies} Through the experiment with bilayer graphene, we have demonstrated the feasibility of using the MLEPI pipeline to study the EPI effect. We aim to further investigate the benefits of incorporating HGNN into HedgeNet. To this end, we conducted an ablation study, comparing the results of HedgeNet and homogeneous GNN (namely, MACE and DeepH-E3) on the bilayer graphene system.

\begin{table}[htbp]
  \caption{Comparison of the mean absolute errors (MAEs) between MACE and HedgeNet-phonon trained on bilayer graphene dataset (energies in meV per atom and forces in meV/\AA). The lowest values are highlighted in \textbf{bold}.}
  \centering
  \resizebox{\linewidth}{!}{
  \begin{tabular}{lrrr}
    \toprule
    ~ & MACE (2 layers) & MACE (4 layers) & HedgeNet-phonon\\
    \midrule
    Energy & 0.1 & 0.1 & \textbf{0.0639} \\
    Force & 9.5 & 9.2 & \textbf{8.686} \\
    \bottomrule
  \end{tabular}}
  \label{tab:phonon-graphene}
\end{table}

\begin{table*}[htbp]
  \caption{Comparison of the MAEs between several benchmark models and HedgeNet-electron, trained on the bilayer graphene dataset (units: meV). The benchmark dataset consists of the test set split from the bilayer graphene dataset and an additional twisted bilayer graphene dataset. The lowest values are highlighted in \textbf{bold}.}
  \centering
  \resizebox{\linewidth}{!}{
  \begin{tabular}{lrrrrr}
    \toprule
    ~ & DeepH & DeepH-E3 (reported) & DeepH-E3 (3 layers) & DeepH-E3 (6 layers) & HedgeNet-electron\\
    \midrule
    Bilayer graphene & 2.1 & 0.40 & 0.395 & 0.310 & \textbf{0.303} \\
    Twisted bilayer graphene & - & - & 0.285 & 0.239 & \textbf{0.206} \\
    \bottomrule
  \end{tabular}}
  \label{tab:electron-graphene}
\end{table*}

As shown in Table~\ref{tab:phonon-graphene}, we trained the MACE model~\cite{batatia2022mace} using the same settings as in the training dataset of the HedgeNet-Phonon in our bilayer graphene experiment. 
To ensure a more fair comparison, we trained both MACE (2 layers) and MACE (4 layers) models. The latter model has twice the number of layers in MACE (2 layers); therefore, its model size is similar to that of HedgeNet-Phonon. It can be seen that the error of MACE (4 layers) on the test set is slightly reduced compared to MACE (2 layers), while the error of HedgeNet-Phonon is the lowest. This indicates that introducing a heterogeneous graph into the phonon network helps to describe the interlayer and intralayer interactions of multilayer materials.

\begin{figure}[htbp]
\includegraphics[width=0.5\textwidth]{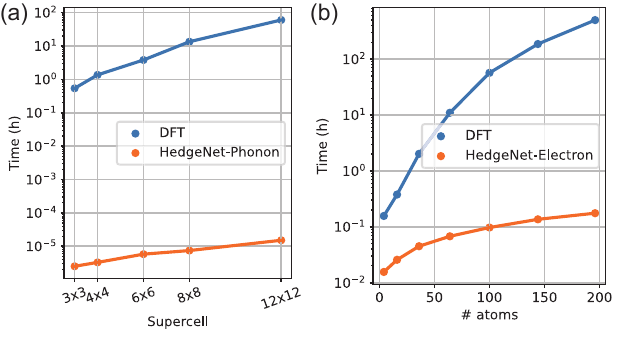}
  \centering
  \caption{Computation time for (a) HedgeNet-Phonon inference and (b) HedgeNet-Electron inference applied to bilayer graphene as a function of system size, with DFT calculation time included for reference.
  }
  \label{fig:6}
\end{figure}

Similarly, we also conducted a comparison for the electronic network. In Table~\ref{tab:electron-graphene}, we compared DeepH~\cite{li2022deep}, DeepH-E3~\cite{gong2023general}, and HedgeNet-Electron on bilayer graphene. The data for DeepH and DeepH-E3 (reported) in the table was extracted from the respective literature, and DeepH-E3 (3 layers) represents our result reproduced according to the DeepH-E3 literature. Our reproduced result matches that in Ref.~\cite{gong2023general}. Like the phonon network, in order to make a fair comparison, we doubled the number of layers of DeepH-E3, denoted as DeepH-E3 (6 layers), whose model size is close to HedgeNet-Electron with DeepH-E3 as its backbone. DeepH-E3 (6 layers) outperforms DeepH-E3 (3 layers) on the test set, but the error can be further reduced by introducing a heterogeneous graph. In addition, Refs.~\cite{li2022deep} and~\cite{gong2023general} also provided a test set for twisted bilayer graphene, which can further explore the model's generalization ability. Therefore, the related results are also shown in Table~\ref{tab:electron-graphene}, again demonstrating the advantages of HedgeNet-Electron. This suggests that introducing a heterogeneous graph to describe multilayer materials in the electronic network is also beneficial.

\subsection{Efficiency of HedgeNet}
DFT enables accurate simulations of the physical properties of materials, but the high computational cost of calculating EPI effects greatly limits the range of systems that can currently be studied. ML methods provide an alternative solution, allowing for more efficient simulation of EPI effects in larger systems. Figures ~\ref{fig:6}(a) and \ref{fig:6}(b) respectively show the time required for HedgeNet inference in the phonon part and electron part as the system size increases, with the DFT calculation time included for reference. The inference for the HedgeNet model was performed on a single 80GB Nvidia Tesla A100 GPU, while the DFT calculations were completed on a node equipped with a 24-core Intel(R) Xeon(R) Silver 4114 CPU @ 2.20GHz. The calculation time for DFT increases with a trend of $N^3$, where $N$ denotes the size of the system, while ML inference grows linearly. Remarkably, ML evaluates individual configurations orders of magnitude faster than DFT—a disparity not attributable to GPU acceleration, which typically improves DFT performance within the same order of magnitude. Note that Fig.~\ref{fig:6}(b) reflects the cost for a single configuration; in practice, converging MC sampling for EPI effects requires tens to hundreds of samples, multiplying computational expense. Thus, ML offers a highly promising approach for investigating EPI effects.

\subsection{Related works and computational complexity}
In heterogeneous relational message passing network (HermNet)~\cite{wang2022heterogeneous}, a concept similar to HedgeNet is proposed, utilizing HGNNs to study material systems. It uses interactions between element pairs to function. It provides three variants: the heterogeneous pair networks (HPNet), focusing on two-body interactions; and the heterogeneous vertex networks (HVNet) and heterogeneous triadic networks (HTNet), focusing on three-body interactions. Their computational complexities are $\mathcal{O}(N_e^2)$, $\mathcal{O}(N_e)$, and $\mathcal{O}(N_e^3)$, respectively, where $N_e$ is the number of element types. 
However, HermNet only considers a single type of subnetwork and encounters challenges when dealing with multilayer materials, as the interactions within and between layers are not the same. HedgeNet addresses this issue more effectively by modeling the interactions within and between layers separately; therefore, HedgeNet has a lower computational complexity and maintains a constant $\mathcal{O}(1)$ complexity.

\section{Conclusion}
Understanding EPIs in multilayer materials enables the creation of advanced materials and devices with tailored properties, enhancing mobility, stability, and overall performance. In this work, we introduce HedgeNet, a HGNN-based architecture that surpasses off‑the‑shelf networks by precisely modeling intra‑ and inter‑layer interactions in multilayer materials, accelerating the study of EPI effects. HedgeNet employs finite difference to model EPIs, naturally capturing higher-order effects inaccessible to most ML methods restricted to lowest-order interactions.
Leveraging our ML methods to bypass DFT computations enables efficient exploration of larger, previously inaccessible material systems.
Experiments on bilayer graphene validate the complete workflow, and ablation studies consistently confirm HedgeNet’s superior performance in both phonon and electronic networks.

\section*{Acknowledgments}
This work was supported by the Quantum Science and Technology-National Science and Technology Major Project (Grant 2023ZD0300500), the Basic Science Center Project of NSFC (Grant No. 52388201), the National Key Basic Research and Development Program of China (Grant No. 2024YFA1409100), and the Beijing Advanced Innovation Center for Future Chip (ICFC).

\section*{AUTHOR CONTRIBUTIONS}
Z.W., W.D. and Z.L. conceived the idea and designed the research. Z.L. performed DFT calculations. Z.W. implemented the codes, trained the models, and conducted all experiments. 
Z.L. supervised the work. All authors discussed the results and were involved in the writing of the manuscript.

\section*{Competing interests}
The authors have no conflicts to disclose.


\begin{thebibliography}{73}%
\makeatletter
\providecommand \@ifxundefined [1]{%
 \@ifx{#1\undefined}
}%
\providecommand \@ifnum [1]{%
 \ifnum #1\expandafter \@firstoftwo
 \else \expandafter \@secondoftwo
 \fi
}%
\providecommand \@ifx [1]{%
 \ifx #1\expandafter \@firstoftwo
 \else \expandafter \@secondoftwo
 \fi
}%
\providecommand \natexlab [1]{#1}%
\providecommand \enquote  [1]{``#1''}%
\providecommand \bibnamefont  [1]{#1}%
\providecommand \bibfnamefont [1]{#1}%
\providecommand \citenamefont [1]{#1}%
\providecommand \href@noop [0]{\@secondoftwo}%
\providecommand \href [0]{\begingroup \@sanitize@url \@href}%
\providecommand \@href[1]{\@@startlink{#1}\@@href}%
\providecommand \@@href[1]{\endgroup#1\@@endlink}%
\providecommand \@sanitize@url [0]{\catcode `\\12\catcode `\$12\catcode
  `\&12\catcode `\#12\catcode `\^12\catcode `\_12\catcode `\%12\relax}%
\providecommand \@@startlink[1]{}%
\providecommand \@@endlink[0]{}%
\providecommand \url  [0]{\begingroup\@sanitize@url \@url }%
\providecommand \@url [1]{\endgroup\@href {#1}{\urlprefix }}%
\providecommand \urlprefix  [0]{URL }%
\providecommand \Eprint [0]{\href }%
\providecommand \doibase [0]{https://doi.org/}%
\providecommand \selectlanguage [0]{\@gobble}%
\providecommand \bibinfo  [0]{\@secondoftwo}%
\providecommand \bibfield  [0]{\@secondoftwo}%
\providecommand \translation [1]{[#1]}%
\providecommand \BibitemOpen [0]{}%
\providecommand \bibitemStop [0]{}%
\providecommand \bibitemNoStop [0]{.\EOS\space}%
\providecommand \EOS [0]{\spacefactor3000\relax}%
\providecommand \BibitemShut  [1]{\csname bibitem#1\endcsname}%
\let\auto@bib@innerbib\@empty
\bibitem [{\citenamefont {Giustino}(2017)}]{giustino2017electron}%
  \BibitemOpen
  \bibfield  {author} {\bibinfo {author} {\bibfnamefont {F.}~\bibnamefont
  {Giustino}},\ }\href@noop {} {\bibfield  {journal} {\bibinfo  {journal} {Rev.
  Mod. Phys.}\ }\textbf {\bibinfo {volume} {89}},\ \bibinfo {pages} {015003}
  (\bibinfo {year} {2017})}\BibitemShut {NoStop}%
\bibitem [{\citenamefont {Monserrat}(2018)}]{monserrat2018electron}%
  \BibitemOpen
  \bibfield  {author} {\bibinfo {author} {\bibfnamefont {B.}~\bibnamefont
  {Monserrat}},\ }\href@noop {} {\bibfield  {journal} {\bibinfo  {journal} {J.
  Phys.: Condens.Matter}\ }\textbf {\bibinfo {volume} {30}},\ \bibinfo {pages}
  {083001} (\bibinfo {year} {2018})}\BibitemShut {NoStop}%
\bibitem [{\citenamefont {Antonius}\ \emph {et~al.}(2014)\citenamefont
  {Antonius}, \citenamefont {Poncé}, \citenamefont {Boulanger}, \citenamefont
  {Côté},\ and\ \citenamefont {Gonze}}]{antonius_many-body_2014}%
  \BibitemOpen
  \bibfield  {author} {\bibinfo {author} {\bibfnamefont {G.}~\bibnamefont
  {Antonius}}, \bibinfo {author} {\bibfnamefont {S.}~\bibnamefont {Poncé}},
  \bibinfo {author} {\bibfnamefont {P.}~\bibnamefont {Boulanger}}, \bibinfo
  {author} {\bibfnamefont {M.}~\bibnamefont {Côté}},\ and\ \bibinfo {author}
  {\bibfnamefont {X.}~\bibnamefont {Gonze}},\ }\href
  {https://doi.org/10.1103/PhysRevLett.112.215501} {\bibfield  {journal}
  {\bibinfo  {journal} {Phys. Rev. Lett.}\ }\textbf {\bibinfo {volume} {112}},\
  \bibinfo {pages} {215501} (\bibinfo {year} {2014})}\BibitemShut {NoStop}%
\bibitem [{\citenamefont
  {Monserrat}(2016{\natexlab{a}})}]{monserrat_correlation_2016}%
  \BibitemOpen
  \bibfield  {author} {\bibinfo {author} {\bibfnamefont {B.}~\bibnamefont
  {Monserrat}},\ }\href {https://doi.org/10.1103/PhysRevB.93.100301} {\bibfield
   {journal} {\bibinfo  {journal} {Phys. Rev. B}\ }\textbf {\bibinfo {volume}
  {93}},\ \bibinfo {pages} {100301} (\bibinfo {year}
  {2016}{\natexlab{a}})}\BibitemShut {NoStop}%
\bibitem [{\citenamefont {Lee}\ \emph {et~al.}(2020)\citenamefont {Lee},
  \citenamefont {Zhou}, \citenamefont {Chen},\ and\ \citenamefont
  {Bernardi}}]{lee2020ab}%
  \BibitemOpen
  \bibfield  {author} {\bibinfo {author} {\bibfnamefont {N.-E.}\ \bibnamefont
  {Lee}}, \bibinfo {author} {\bibfnamefont {J.-J.}\ \bibnamefont {Zhou}},
  \bibinfo {author} {\bibfnamefont {H.-Y.}\ \bibnamefont {Chen}},\ and\
  \bibinfo {author} {\bibfnamefont {M.}~\bibnamefont {Bernardi}},\ }\href@noop
  {} {\bibfield  {journal} {\bibinfo  {journal} {Nat. Commun.}\ }\textbf
  {\bibinfo {volume} {11}},\ \bibinfo {pages} {1607} (\bibinfo {year}
  {2020})}\BibitemShut {NoStop}%
\bibitem [{\citenamefont {Alldredge}\ and\ \citenamefont
  {Blatt}(1967)}]{alldredge1967role}%
  \BibitemOpen
  \bibfield  {author} {\bibinfo {author} {\bibfnamefont {G.~P.}\ \bibnamefont
  {Alldredge}}\ and\ \bibinfo {author} {\bibfnamefont {F.}~\bibnamefont
  {Blatt}},\ }\href@noop {} {\bibfield  {journal} {\bibinfo  {journal} {Ann.
  Phys.}\ }\textbf {\bibinfo {volume} {45}},\ \bibinfo {pages} {191} (\bibinfo
  {year} {1967})}\BibitemShut {NoStop}%
\bibitem [{\citenamefont {Thorbergsson}\ and\ \citenamefont
  {Sak}(1986)}]{thorbergsson1986mobility}%
  \BibitemOpen
  \bibfield  {author} {\bibinfo {author} {\bibfnamefont {G.~I.}\ \bibnamefont
  {Thorbergsson}}\ and\ \bibinfo {author} {\bibfnamefont {J.}~\bibnamefont
  {Sak}},\ }\href@noop {} {\bibfield  {journal} {\bibinfo  {journal} {Phys.
  Lett. A}\ }\textbf {\bibinfo {volume} {116}},\ \bibinfo {pages} {325}
  (\bibinfo {year} {1986})}\BibitemShut {NoStop}%
\bibitem [{\citenamefont {Sher}\ and\ \citenamefont
  {Thornber}(1967)}]{sher1967resonant}%
  \BibitemOpen
  \bibfield  {author} {\bibinfo {author} {\bibfnamefont {A.}~\bibnamefont
  {Sher}}\ and\ \bibinfo {author} {\bibfnamefont {K.}~\bibnamefont
  {Thornber}},\ }\href@noop {} {\bibfield  {journal} {\bibinfo  {journal}
  {Appl. Phys. Lett.}\ }\textbf {\bibinfo {volume} {11}},\ \bibinfo {pages} {3}
  (\bibinfo {year} {1967})}\BibitemShut {NoStop}%
\bibitem [{\citenamefont {Woods}\ and\ \citenamefont
  {Mahan}(1998)}]{woods1998nonlinear}%
  \BibitemOpen
  \bibfield  {author} {\bibinfo {author} {\bibfnamefont {L.}~\bibnamefont
  {Woods}}\ and\ \bibinfo {author} {\bibfnamefont {G.}~\bibnamefont {Mahan}},\
  }\href@noop {} {\bibfield  {journal} {\bibinfo  {journal} {Phys. Rev. B}\
  }\textbf {\bibinfo {volume} {57}},\ \bibinfo {pages} {7679} (\bibinfo {year}
  {1998})}\BibitemShut {NoStop}%
\bibitem [{\citenamefont {Wang}\ and\ \citenamefont
  {Mahan}(1989)}]{wang1989low}%
  \BibitemOpen
  \bibfield  {author} {\bibinfo {author} {\bibfnamefont {Z.}~\bibnamefont
  {Wang}}\ and\ \bibinfo {author} {\bibfnamefont {G.}~\bibnamefont {Mahan}},\
  }\href@noop {} {\bibfield  {journal} {\bibinfo  {journal} {Phys. Rev. B}\
  }\textbf {\bibinfo {volume} {39}},\ \bibinfo {pages} {10753} (\bibinfo {year}
  {1989})}\BibitemShut {NoStop}%
\bibitem [{\citenamefont {Story}\ \emph {et~al.}(2014)\citenamefont {Story},
  \citenamefont {Kas}, \citenamefont {Vila}, \citenamefont {Verstraete},\ and\
  \citenamefont {Rehr}}]{story2014cumulant}%
  \BibitemOpen
  \bibfield  {author} {\bibinfo {author} {\bibfnamefont {S.}~\bibnamefont
  {Story}}, \bibinfo {author} {\bibfnamefont {J.}~\bibnamefont {Kas}}, \bibinfo
  {author} {\bibfnamefont {F.}~\bibnamefont {Vila}}, \bibinfo {author}
  {\bibfnamefont {M.}~\bibnamefont {Verstraete}},\ and\ \bibinfo {author}
  {\bibfnamefont {J.}~\bibnamefont {Rehr}},\ }\href@noop {} {\bibfield
  {journal} {\bibinfo  {journal} {Phys. Rev. B}\ }\textbf {\bibinfo {volume}
  {90}},\ \bibinfo {pages} {195135} (\bibinfo {year} {2014})}\BibitemShut
  {NoStop}%
\bibitem [{\citenamefont {Nery}\ \emph {et~al.}(2018)\citenamefont {Nery},
  \citenamefont {Allen}, \citenamefont {Antonius}, \citenamefont {Reining},
  \citenamefont {Miglio},\ and\ \citenamefont
  {Gonze}}]{nery2018quasiparticles}%
  \BibitemOpen
  \bibfield  {author} {\bibinfo {author} {\bibfnamefont {J.~P.}\ \bibnamefont
  {Nery}}, \bibinfo {author} {\bibfnamefont {P.~B.}\ \bibnamefont {Allen}},
  \bibinfo {author} {\bibfnamefont {G.}~\bibnamefont {Antonius}}, \bibinfo
  {author} {\bibfnamefont {L.}~\bibnamefont {Reining}}, \bibinfo {author}
  {\bibfnamefont {A.}~\bibnamefont {Miglio}},\ and\ \bibinfo {author}
  {\bibfnamefont {X.}~\bibnamefont {Gonze}},\ }\href@noop {} {\bibfield
  {journal} {\bibinfo  {journal} {Phys. Rev. B}\ }\textbf {\bibinfo {volume}
  {97}},\ \bibinfo {pages} {115145} (\bibinfo {year} {2018})}\BibitemShut
  {NoStop}%
\bibitem [{\citenamefont {Zhou}\ and\ \citenamefont
  {Bernardi}(2019)}]{zhou2019predicting}%
  \BibitemOpen
  \bibfield  {author} {\bibinfo {author} {\bibfnamefont {J.-J.}\ \bibnamefont
  {Zhou}}\ and\ \bibinfo {author} {\bibfnamefont {M.}~\bibnamefont
  {Bernardi}},\ }\href@noop {} {\bibfield  {journal} {\bibinfo  {journal}
  {Phys. Rev. Res.}\ }\textbf {\bibinfo {volume} {1}},\ \bibinfo {pages}
  {033138} (\bibinfo {year} {2019})}\BibitemShut {NoStop}%
\bibitem [{\citenamefont {Allen}\ and\ \citenamefont
  {Heine}(1976)}]{allen_theory_1976}%
  \BibitemOpen
  \bibfield  {author} {\bibinfo {author} {\bibfnamefont {P.~B.}\ \bibnamefont
  {Allen}}\ and\ \bibinfo {author} {\bibfnamefont {V.}~\bibnamefont {Heine}},\
  }\href {https://doi.org/10.1088/0022-3719/9/12/013} {\bibfield  {journal}
  {\bibinfo  {journal} {J. Phys. C: Solid State Phys.}\ }\textbf {\bibinfo
  {volume} {9}},\ \bibinfo {pages} {2305} (\bibinfo {year} {1976})}\BibitemShut
  {NoStop}%
\bibitem [{\citenamefont {Lin}\ \emph {et~al.}(2022)\citenamefont {Lin},
  \citenamefont {Liu}, \citenamefont {Wang}, \citenamefont {Xu}, \citenamefont
  {Chen}, \citenamefont {Duan},\ and\ \citenamefont
  {Monserrat}}]{lin_phonon-limited_2022}%
  \BibitemOpen
  \bibfield  {author} {\bibinfo {author} {\bibfnamefont {Z.}~\bibnamefont
  {Lin}}, \bibinfo {author} {\bibfnamefont {Y.}~\bibnamefont {Liu}}, \bibinfo
  {author} {\bibfnamefont {Z.}~\bibnamefont {Wang}}, \bibinfo {author}
  {\bibfnamefont {S.}~\bibnamefont {Xu}}, \bibinfo {author} {\bibfnamefont
  {S.}~\bibnamefont {Chen}}, \bibinfo {author} {\bibfnamefont {W.}~\bibnamefont
  {Duan}},\ and\ \bibinfo {author} {\bibfnamefont {B.}~\bibnamefont
  {Monserrat}},\ }\href {https://doi.org/10.1103/PhysRevLett.129.027401}
  {\bibfield  {journal} {\bibinfo  {journal} {Phys. Rev. Lett.}\ }\textbf
  {\bibinfo {volume} {129}},\ \bibinfo {pages} {027401} (\bibinfo {year}
  {2022})}\BibitemShut {NoStop}%
\bibitem [{\citenamefont {Lloyd-Williams}\ and\ \citenamefont
  {Monserrat}(2015)}]{lloyd-williams_lattice_2015}%
  \BibitemOpen
  \bibfield  {author} {\bibinfo {author} {\bibfnamefont {J.~H.}\ \bibnamefont
  {Lloyd-Williams}}\ and\ \bibinfo {author} {\bibfnamefont {B.}~\bibnamefont
  {Monserrat}},\ }\href {https://doi.org/10.1103/PhysRevB.92.184301} {\bibfield
   {journal} {\bibinfo  {journal} {Phys. Rev. B}\ }\textbf {\bibinfo {volume}
  {92}},\ \bibinfo {pages} {184301} (\bibinfo {year} {2015})}\BibitemShut
  {NoStop}%
\bibitem [{\citenamefont {Chen}\ \emph {et~al.}(2022)\citenamefont {Chen},
  \citenamefont {Salzbrenner},\ and\ \citenamefont
  {Monserrat}}]{chen_nonuniform_2022}%
  \BibitemOpen
  \bibfield  {author} {\bibinfo {author} {\bibfnamefont {S.}~\bibnamefont
  {Chen}}, \bibinfo {author} {\bibfnamefont {P.~T.}\ \bibnamefont
  {Salzbrenner}},\ and\ \bibinfo {author} {\bibfnamefont {B.}~\bibnamefont
  {Monserrat}},\ }\href {https://doi.org/10.1103/PhysRevB.106.155102}
  {\bibfield  {journal} {\bibinfo  {journal} {Phys. Rev. B}\ }\textbf {\bibinfo
  {volume} {106}},\ \bibinfo {pages} {155102} (\bibinfo {year}
  {2022})}\BibitemShut {NoStop}%
\bibitem [{\citenamefont {Patrick}\ and\ \citenamefont
  {Giustino}(2013)}]{patrick_quantum_2013}%
  \BibitemOpen
  \bibfield  {author} {\bibinfo {author} {\bibfnamefont {C.~E.}\ \bibnamefont
  {Patrick}}\ and\ \bibinfo {author} {\bibfnamefont {F.}~\bibnamefont
  {Giustino}},\ }\href {https://doi.org/10.1038/ncomms3006} {\bibfield
  {journal} {\bibinfo  {journal} {Nat Commun}\ }\textbf {\bibinfo {volume}
  {4}},\ \bibinfo {pages} {2006} (\bibinfo {year} {2013})}\BibitemShut
  {NoStop}%
\bibitem [{\citenamefont
  {Monserrat}(2016{\natexlab{b}})}]{monserrat_vibrational_2016}%
  \BibitemOpen
  \bibfield  {author} {\bibinfo {author} {\bibfnamefont {B.}~\bibnamefont
  {Monserrat}},\ }\href {https://doi.org/10.1103/PhysRevB.93.014302} {\bibfield
   {journal} {\bibinfo  {journal} {Phys. Rev. B}\ }\textbf {\bibinfo {volume}
  {93}},\ \bibinfo {pages} {014302} (\bibinfo {year}
  {2016}{\natexlab{b}})}\BibitemShut {NoStop}%
\bibitem [{\citenamefont {Zacharias}\ and\ \citenamefont
  {Giustino}(2016)}]{zacharias_one-shot_2016}%
  \BibitemOpen
  \bibfield  {author} {\bibinfo {author} {\bibfnamefont {M.}~\bibnamefont
  {Zacharias}}\ and\ \bibinfo {author} {\bibfnamefont {F.}~\bibnamefont
  {Giustino}},\ }\href {https://doi.org/10.1103/PhysRevB.94.075125} {\bibfield
  {journal} {\bibinfo  {journal} {Phys. Rev. B}\ }\textbf {\bibinfo {volume}
  {94}},\ \bibinfo {pages} {075125} (\bibinfo {year} {2016})}\BibitemShut
  {NoStop}%
\bibitem [{\citenamefont
  {Monserrat}(2016{\natexlab{c}})}]{monserrat_correlation_2016-1}%
  \BibitemOpen
  \bibfield  {author} {\bibinfo {author} {\bibfnamefont {B.}~\bibnamefont
  {Monserrat}},\ }\href {https://doi.org/10.1103/PhysRevB.93.100301} {\bibfield
   {journal} {\bibinfo  {journal} {Phys. Rev. B}\ }\textbf {\bibinfo {volume}
  {93}},\ \bibinfo {pages} {100301} (\bibinfo {year}
  {2016}{\natexlab{c}})}\BibitemShut {NoStop}%
\bibitem [{\citenamefont {Jordan}\ and\ \citenamefont
  {Mitchell}(2015)}]{jordan2015machine}%
  \BibitemOpen
  \bibfield  {author} {\bibinfo {author} {\bibfnamefont {M.~I.}\ \bibnamefont
  {Jordan}}\ and\ \bibinfo {author} {\bibfnamefont {T.~M.}\ \bibnamefont
  {Mitchell}},\ }\href@noop {} {\bibfield  {journal} {\bibinfo  {journal}
  {Science}\ }\textbf {\bibinfo {volume} {349}},\ \bibinfo {pages} {255}
  (\bibinfo {year} {2015})}\BibitemShut {NoStop}%
\bibitem [{\citenamefont {Wang}\ \emph
  {et~al.}(2023{\natexlab{a}})\citenamefont {Wang}, \citenamefont {Fu},
  \citenamefont {Du}, \citenamefont {Gao}, \citenamefont {Huang}, \citenamefont
  {Liu}, \citenamefont {Chandak}, \citenamefont {Liu}, \citenamefont
  {Van~Katwyk}, \citenamefont {Deac} \emph {et~al.}}]{wang2023scientific}%
  \BibitemOpen
  \bibfield  {author} {\bibinfo {author} {\bibfnamefont {H.}~\bibnamefont
  {Wang}}, \bibinfo {author} {\bibfnamefont {T.}~\bibnamefont {Fu}}, \bibinfo
  {author} {\bibfnamefont {Y.}~\bibnamefont {Du}}, \bibinfo {author}
  {\bibfnamefont {W.}~\bibnamefont {Gao}}, \bibinfo {author} {\bibfnamefont
  {K.}~\bibnamefont {Huang}}, \bibinfo {author} {\bibfnamefont
  {Z.}~\bibnamefont {Liu}}, \bibinfo {author} {\bibfnamefont {P.}~\bibnamefont
  {Chandak}}, \bibinfo {author} {\bibfnamefont {S.}~\bibnamefont {Liu}},
  \bibinfo {author} {\bibfnamefont {P.}~\bibnamefont {Van~Katwyk}}, \bibinfo
  {author} {\bibfnamefont {A.}~\bibnamefont {Deac}}, \emph {et~al.},\
  }\href@noop {} {\bibfield  {journal} {\bibinfo  {journal} {Nature}\ }\textbf
  {\bibinfo {volume} {620}},\ \bibinfo {pages} {47} (\bibinfo {year}
  {2023}{\natexlab{a}})}\BibitemShut {NoStop}%
\bibitem [{\citenamefont {Behler}\ and\ \citenamefont
  {Parrinello}(2007)}]{behler2007generalized}%
  \BibitemOpen
  \bibfield  {author} {\bibinfo {author} {\bibfnamefont {J.}~\bibnamefont
  {Behler}}\ and\ \bibinfo {author} {\bibfnamefont {M.}~\bibnamefont
  {Parrinello}},\ }\href@noop {} {\bibfield  {journal} {\bibinfo  {journal}
  {Phys. Rev. Lett.}\ }\textbf {\bibinfo {volume} {98}},\ \bibinfo {pages}
  {146401} (\bibinfo {year} {2007})}\BibitemShut {NoStop}%
\bibitem [{\citenamefont {Sch{\"u}tt}\ \emph {et~al.}(2017)\citenamefont
  {Sch{\"u}tt}, \citenamefont {Kindermans}, \citenamefont {Sauceda~Felix},
  \citenamefont {Chmiela}, \citenamefont {Tkatchenko},\ and\ \citenamefont
  {M{\"u}ller}}]{schutt2017schnet}%
  \BibitemOpen
  \bibfield  {author} {\bibinfo {author} {\bibfnamefont {K.}~\bibnamefont
  {Sch{\"u}tt}}, \bibinfo {author} {\bibfnamefont {P.-J.}\ \bibnamefont
  {Kindermans}}, \bibinfo {author} {\bibfnamefont {H.~E.}\ \bibnamefont
  {Sauceda~Felix}}, \bibinfo {author} {\bibfnamefont {S.}~\bibnamefont
  {Chmiela}}, \bibinfo {author} {\bibfnamefont {A.}~\bibnamefont
  {Tkatchenko}},\ and\ \bibinfo {author} {\bibfnamefont {K.-R.}\ \bibnamefont
  {M{\"u}ller}},\ }\href@noop {} {\bibfield  {journal} {\bibinfo  {journal}
  {Advances in neural information processing systems}\ }\textbf {\bibinfo
  {volume} {30}} (\bibinfo {year} {2017})}\BibitemShut {NoStop}%
\bibitem [{\citenamefont {Kindermans}\ and\ \citenamefont
  {M{\"u}ller}(2018)}]{schutt2018schnet}%
  \BibitemOpen
  \bibfield  {author} {\bibinfo {author} {\bibfnamefont {P.-J.}\ \bibnamefont
  {Kindermans}}\ and\ \bibinfo {author} {\bibfnamefont {K.-R.}\ \bibnamefont
  {M{\"u}ller}},\ }\href@noop {} {\bibfield  {journal} {\bibinfo  {journal} {J.
  Chem. Phys.}\ }\textbf {\bibinfo {volume} {148}} (\bibinfo {year}
  {2018})}\BibitemShut {NoStop}%
\bibitem [{\citenamefont {Thomas}\ \emph {et~al.}(2018)\citenamefont {Thomas},
  \citenamefont {Smidt}, \citenamefont {Kearnes}, \citenamefont {Yang},
  \citenamefont {Li}, \citenamefont {Kohlhoff},\ and\ \citenamefont
  {Riley}}]{thomas2018tensor}%
  \BibitemOpen
  \bibfield  {author} {\bibinfo {author} {\bibfnamefont {N.}~\bibnamefont
  {Thomas}}, \bibinfo {author} {\bibfnamefont {T.}~\bibnamefont {Smidt}},
  \bibinfo {author} {\bibfnamefont {S.}~\bibnamefont {Kearnes}}, \bibinfo
  {author} {\bibfnamefont {L.}~\bibnamefont {Yang}}, \bibinfo {author}
  {\bibfnamefont {L.}~\bibnamefont {Li}}, \bibinfo {author} {\bibfnamefont
  {K.}~\bibnamefont {Kohlhoff}},\ and\ \bibinfo {author} {\bibfnamefont
  {P.}~\bibnamefont {Riley}},\ }\href@noop {} {\bibfield  {journal} {\bibinfo
  {journal} {Preprint at \url{http://arxiv.org/abs/1802.08219}}\ } (\bibinfo
  {year} {2018})}\BibitemShut {NoStop}%
\bibitem [{\citenamefont {Coors}\ \emph {et~al.}(2018)\citenamefont {Coors},
  \citenamefont {Condurache},\ and\ \citenamefont
  {Geiger}}]{coors2018spherenet}%
  \BibitemOpen
  \bibfield  {author} {\bibinfo {author} {\bibfnamefont {B.}~\bibnamefont
  {Coors}}, \bibinfo {author} {\bibfnamefont {A.~P.}\ \bibnamefont
  {Condurache}},\ and\ \bibinfo {author} {\bibfnamefont {A.}~\bibnamefont
  {Geiger}},\ }in\ \href@noop {} {\emph {\bibinfo {booktitle} {Proceedings of
  the European conference on computer vision (ECCV)}}}\ (\bibinfo {year}
  {2018})\ pp.\ \bibinfo {pages} {518--533}\BibitemShut {NoStop}%
\bibitem [{\citenamefont {Anderson}\ \emph {et~al.}(2019)\citenamefont
  {Anderson}, \citenamefont {Hy},\ and\ \citenamefont
  {Kondor}}]{anderson2019cormorant}%
  \BibitemOpen
  \bibfield  {author} {\bibinfo {author} {\bibfnamefont {B.}~\bibnamefont
  {Anderson}}, \bibinfo {author} {\bibfnamefont {T.~S.}\ \bibnamefont {Hy}},\
  and\ \bibinfo {author} {\bibfnamefont {R.}~\bibnamefont {Kondor}},\
  }\href@noop {} {\bibfield  {journal} {\bibinfo  {journal} {Advances in neural
  information processing systems}\ }\textbf {\bibinfo {volume} {32}} (\bibinfo
  {year} {2019})}\BibitemShut {NoStop}%
\bibitem [{\citenamefont {Gasteiger}\ \emph {et~al.}(2019)\citenamefont
  {Gasteiger}, \citenamefont {Gro{\ss}},\ and\ \citenamefont
  {G{\"u}nnemann}}]{gasteiger2019directional}%
  \BibitemOpen
  \bibfield  {author} {\bibinfo {author} {\bibfnamefont {J.}~\bibnamefont
  {Gasteiger}}, \bibinfo {author} {\bibfnamefont {J.}~\bibnamefont
  {Gro{\ss}}},\ and\ \bibinfo {author} {\bibfnamefont {S.}~\bibnamefont
  {G{\"u}nnemann}},\ }in\ \href@noop {} {\emph {\bibinfo {booktitle}
  {International Conference on Learning Representations}}}\ (\bibinfo {year}
  {2019})\BibitemShut {NoStop}%
\bibitem [{\citenamefont {Gasteiger}\ \emph {et~al.}(2020)\citenamefont
  {Gasteiger}, \citenamefont {Giri}, \citenamefont {Margraf},\ and\
  \citenamefont {G{\"u}nnemann}}]{gasteiger2020fast}%
  \BibitemOpen
  \bibfield  {author} {\bibinfo {author} {\bibfnamefont {J.}~\bibnamefont
  {Gasteiger}}, \bibinfo {author} {\bibfnamefont {S.}~\bibnamefont {Giri}},
  \bibinfo {author} {\bibfnamefont {J.~T.}\ \bibnamefont {Margraf}},\ and\
  \bibinfo {author} {\bibfnamefont {S.}~\bibnamefont {G{\"u}nnemann}},\
  }\href@noop {} {\bibfield  {journal} {\bibinfo  {journal} {Preprint at
  \url{http://arxiv.org/abs/2011.14115}}\ } (\bibinfo {year}
  {2020})}\BibitemShut {NoStop}%
\bibitem [{\citenamefont {Fuchs}\ \emph {et~al.}(2020)\citenamefont {Fuchs},
  \citenamefont {Worrall}, \citenamefont {Fischer},\ and\ \citenamefont
  {Welling}}]{fuchs2020se}%
  \BibitemOpen
  \bibfield  {author} {\bibinfo {author} {\bibfnamefont {F.}~\bibnamefont
  {Fuchs}}, \bibinfo {author} {\bibfnamefont {D.}~\bibnamefont {Worrall}},
  \bibinfo {author} {\bibfnamefont {V.}~\bibnamefont {Fischer}},\ and\ \bibinfo
  {author} {\bibfnamefont {M.}~\bibnamefont {Welling}},\ }\href@noop {}
  {\bibfield  {journal} {\bibinfo  {journal} {Advances in Neural Information
  Processing Systems}\ }\textbf {\bibinfo {volume} {33}},\ \bibinfo {pages}
  {1970} (\bibinfo {year} {2020})}\BibitemShut {NoStop}%
\bibitem [{\citenamefont {Sch{\"u}tt}\ \emph {et~al.}(2021)\citenamefont
  {Sch{\"u}tt}, \citenamefont {Unke},\ and\ \citenamefont
  {Gastegger}}]{schutt2021equivariant}%
  \BibitemOpen
  \bibfield  {author} {\bibinfo {author} {\bibfnamefont {K.}~\bibnamefont
  {Sch{\"u}tt}}, \bibinfo {author} {\bibfnamefont {O.}~\bibnamefont {Unke}},\
  and\ \bibinfo {author} {\bibfnamefont {M.}~\bibnamefont {Gastegger}},\ }in\
  \href@noop {} {\emph {\bibinfo {booktitle} {International Conference on
  Machine Learning}}}\ (\bibinfo {organization} {PMLR},\ \bibinfo {year}
  {2021})\ pp.\ \bibinfo {pages} {9377--9388}\BibitemShut {NoStop}%
\bibitem [{\citenamefont {Gasteiger}\ \emph {et~al.}(2021)\citenamefont
  {Gasteiger}, \citenamefont {Becker},\ and\ \citenamefont
  {G{\"u}nnemann}}]{gasteiger2021gemnet}%
  \BibitemOpen
  \bibfield  {author} {\bibinfo {author} {\bibfnamefont {J.}~\bibnamefont
  {Gasteiger}}, \bibinfo {author} {\bibfnamefont {F.}~\bibnamefont {Becker}},\
  and\ \bibinfo {author} {\bibfnamefont {S.}~\bibnamefont {G{\"u}nnemann}},\
  }\href@noop {} {\bibfield  {journal} {\bibinfo  {journal} {Advances in Neural
  Information Processing Systems}\ }\textbf {\bibinfo {volume} {34}},\ \bibinfo
  {pages} {6790} (\bibinfo {year} {2021})}\BibitemShut {NoStop}%
\bibitem [{\citenamefont {Th{\"o}lke}\ and\ \citenamefont
  {De~Fabritiis}(2021)}]{tholke2021equivariant}%
  \BibitemOpen
  \bibfield  {author} {\bibinfo {author} {\bibfnamefont {P.}~\bibnamefont
  {Th{\"o}lke}}\ and\ \bibinfo {author} {\bibfnamefont {G.}~\bibnamefont
  {De~Fabritiis}},\ }in\ \href@noop {} {\emph {\bibinfo {booktitle}
  {International Conference on Learning Representations}}}\ (\bibinfo {year}
  {2021})\BibitemShut {NoStop}%
\bibitem [{\citenamefont {Batzner}\ \emph {et~al.}(2022)\citenamefont
  {Batzner}, \citenamefont {Musaelian}, \citenamefont {Sun}, \citenamefont
  {Geiger}, \citenamefont {Mailoa}, \citenamefont {Kornbluth}, \citenamefont
  {Molinari}, \citenamefont {Smidt},\ and\ \citenamefont
  {Kozinsky}}]{batzner2022}%
  \BibitemOpen
  \bibfield  {author} {\bibinfo {author} {\bibfnamefont {S.}~\bibnamefont
  {Batzner}}, \bibinfo {author} {\bibfnamefont {A.}~\bibnamefont {Musaelian}},
  \bibinfo {author} {\bibfnamefont {L.}~\bibnamefont {Sun}}, \bibinfo {author}
  {\bibfnamefont {M.}~\bibnamefont {Geiger}}, \bibinfo {author} {\bibfnamefont
  {J.~P.}\ \bibnamefont {Mailoa}}, \bibinfo {author} {\bibfnamefont
  {M.}~\bibnamefont {Kornbluth}}, \bibinfo {author} {\bibfnamefont
  {N.}~\bibnamefont {Molinari}}, \bibinfo {author} {\bibfnamefont {T.~E.}\
  \bibnamefont {Smidt}},\ and\ \bibinfo {author} {\bibfnamefont
  {B.}~\bibnamefont {Kozinsky}},\ }\href@noop {} {\bibfield  {journal}
  {\bibinfo  {journal} {Nat. Commun.}\ }\textbf {\bibinfo {volume} {13}},\
  \bibinfo {pages} {2453} (\bibinfo {year} {2022})}\BibitemShut {NoStop}%
\bibitem [{\citenamefont {Musaelian}\ \emph {et~al.}(2023)\citenamefont
  {Musaelian}, \citenamefont {Batzner}, \citenamefont {Johansson},
  \citenamefont {Sun}, \citenamefont {Owen}, \citenamefont {Kornbluth},\ and\
  \citenamefont {Kozinsky}}]{musaelian2023learning}%
  \BibitemOpen
  \bibfield  {author} {\bibinfo {author} {\bibfnamefont {A.}~\bibnamefont
  {Musaelian}}, \bibinfo {author} {\bibfnamefont {S.}~\bibnamefont {Batzner}},
  \bibinfo {author} {\bibfnamefont {A.}~\bibnamefont {Johansson}}, \bibinfo
  {author} {\bibfnamefont {L.}~\bibnamefont {Sun}}, \bibinfo {author}
  {\bibfnamefont {C.~J.}\ \bibnamefont {Owen}}, \bibinfo {author}
  {\bibfnamefont {M.}~\bibnamefont {Kornbluth}},\ and\ \bibinfo {author}
  {\bibfnamefont {B.}~\bibnamefont {Kozinsky}},\ }\href@noop {} {\bibfield
  {journal} {\bibinfo  {journal} {Nat. Commun.}\ }\textbf {\bibinfo {volume}
  {14}},\ \bibinfo {pages} {579} (\bibinfo {year} {2023})}\BibitemShut
  {NoStop}%
\bibitem [{\citenamefont {Liao}\ and\ \citenamefont
  {Smidt}(2022)}]{liao2022equiformer}%
  \BibitemOpen
  \bibfield  {author} {\bibinfo {author} {\bibfnamefont {Y.-L.}\ \bibnamefont
  {Liao}}\ and\ \bibinfo {author} {\bibfnamefont {T.}~\bibnamefont {Smidt}},\
  }in\ \href@noop {} {\emph {\bibinfo {booktitle} {The Eleventh International
  Conference on Learning Representations}}}\ (\bibinfo {year}
  {2022})\BibitemShut {NoStop}%
\bibitem [{\citenamefont {Batatia}\ \emph {et~al.}(2022)\citenamefont
  {Batatia}, \citenamefont {Kovacs}, \citenamefont {Simm}, \citenamefont
  {Ortner},\ and\ \citenamefont {Cs{\'a}nyi}}]{batatia2022mace}%
  \BibitemOpen
  \bibfield  {author} {\bibinfo {author} {\bibfnamefont {I.}~\bibnamefont
  {Batatia}}, \bibinfo {author} {\bibfnamefont {D.~P.}\ \bibnamefont {Kovacs}},
  \bibinfo {author} {\bibfnamefont {G.}~\bibnamefont {Simm}}, \bibinfo {author}
  {\bibfnamefont {C.}~\bibnamefont {Ortner}},\ and\ \bibinfo {author}
  {\bibfnamefont {G.}~\bibnamefont {Cs{\'a}nyi}},\ }\href@noop {} {\bibfield
  {journal} {\bibinfo  {journal} {Advances in Neural Information Processing
  Systems}\ }\textbf {\bibinfo {volume} {35}},\ \bibinfo {pages} {11423}
  (\bibinfo {year} {2022})}\BibitemShut {NoStop}%
\bibitem [{\citenamefont {Wang}\ \emph
  {et~al.}(2022{\natexlab{a}})\citenamefont {Wang}, \citenamefont {Liu},
  \citenamefont {Lin}, \citenamefont {Liu},\ and\ \citenamefont
  {Ji}}]{wang2022comenet}%
  \BibitemOpen
  \bibfield  {author} {\bibinfo {author} {\bibfnamefont {L.}~\bibnamefont
  {Wang}}, \bibinfo {author} {\bibfnamefont {Y.}~\bibnamefont {Liu}}, \bibinfo
  {author} {\bibfnamefont {Y.}~\bibnamefont {Lin}}, \bibinfo {author}
  {\bibfnamefont {H.}~\bibnamefont {Liu}},\ and\ \bibinfo {author}
  {\bibfnamefont {S.}~\bibnamefont {Ji}},\ }in\ \href@noop {} {\emph {\bibinfo
  {booktitle} {Advances in Neural Information Processing Systems}}},\ \bibinfo
  {editor} {edited by\ \bibinfo {editor} {\bibfnamefont {A.~H.}\ \bibnamefont
  {Oh}}, \bibinfo {editor} {\bibfnamefont {A.}~\bibnamefont {Agarwal}},
  \bibinfo {editor} {\bibfnamefont {D.}~\bibnamefont {Belgrave}},\ and\
  \bibinfo {editor} {\bibfnamefont {K.}~\bibnamefont {Cho}}}\ (\bibinfo {year}
  {2022})\BibitemShut {NoStop}%
\bibitem [{\citenamefont {Wang}\ \emph
  {et~al.}(2023{\natexlab{b}})\citenamefont {Wang}, \citenamefont {Liu},
  \citenamefont {Zhou}, \citenamefont {Wang},\ and\ \citenamefont
  {Shao}}]{wang2023efficiently}%
  \BibitemOpen
  \bibfield  {author} {\bibinfo {author} {\bibfnamefont {Z.}~\bibnamefont
  {Wang}}, \bibinfo {author} {\bibfnamefont {G.}~\bibnamefont {Liu}}, \bibinfo
  {author} {\bibfnamefont {Y.}~\bibnamefont {Zhou}}, \bibinfo {author}
  {\bibfnamefont {T.}~\bibnamefont {Wang}},\ and\ \bibinfo {author}
  {\bibfnamefont {B.}~\bibnamefont {Shao}},\ }in\ \href@noop {} {\emph
  {\bibinfo {booktitle} {Thirty-seventh Conference on Neural Information
  Processing Systems}}}\ (\bibinfo {year} {2023})\BibitemShut {NoStop}%
\bibitem [{\citenamefont {Liao}\ \emph {et~al.}()\citenamefont {Liao},
  \citenamefont {Wood}, \citenamefont {Das},\ and\ \citenamefont
  {Smidt}}]{liao2023equiformerv2}%
  \BibitemOpen
  \bibfield  {author} {\bibinfo {author} {\bibfnamefont {Y.-L.}\ \bibnamefont
  {Liao}}, \bibinfo {author} {\bibfnamefont {B.~M.}\ \bibnamefont {Wood}},
  \bibinfo {author} {\bibfnamefont {A.}~\bibnamefont {Das}},\ and\ \bibinfo
  {author} {\bibfnamefont {T.}~\bibnamefont {Smidt}},\ }in\ \href@noop {}
  {\emph {\bibinfo {booktitle} {The Twelfth International Conference on
  Learning Representations}}}\BibitemShut {NoStop}%
\bibitem [{\citenamefont {Wang}\ \emph
  {et~al.}(2024{\natexlab{a}})\citenamefont {Wang}, \citenamefont {Wang},
  \citenamefont {Li}, \citenamefont {He}, \citenamefont {Li}, \citenamefont
  {Wang}, \citenamefont {Zheng}, \citenamefont {Shao},\ and\ \citenamefont
  {Liu}}]{wang2024enhancing}%
  \BibitemOpen
  \bibfield  {author} {\bibinfo {author} {\bibfnamefont {Y.}~\bibnamefont
  {Wang}}, \bibinfo {author} {\bibfnamefont {T.}~\bibnamefont {Wang}}, \bibinfo
  {author} {\bibfnamefont {S.}~\bibnamefont {Li}}, \bibinfo {author}
  {\bibfnamefont {X.}~\bibnamefont {He}}, \bibinfo {author} {\bibfnamefont
  {M.}~\bibnamefont {Li}}, \bibinfo {author} {\bibfnamefont {Z.}~\bibnamefont
  {Wang}}, \bibinfo {author} {\bibfnamefont {N.}~\bibnamefont {Zheng}},
  \bibinfo {author} {\bibfnamefont {B.}~\bibnamefont {Shao}},\ and\ \bibinfo
  {author} {\bibfnamefont {T.-Y.}\ \bibnamefont {Liu}},\ }\href@noop {}
  {\bibfield  {journal} {\bibinfo  {journal} {Nat. Commun.}\ }\textbf {\bibinfo
  {volume} {15}},\ \bibinfo {pages} {313} (\bibinfo {year}
  {2024}{\natexlab{a}})}\BibitemShut {NoStop}%
\bibitem [{\citenamefont {Li}\ \emph {et~al.}(2024{\natexlab{a}})\citenamefont
  {Li}, \citenamefont {Wang}, \citenamefont {Huang}, \citenamefont {Yang},
  \citenamefont {Wei}, \citenamefont {Zhang}, \citenamefont {Wang},
  \citenamefont {Wang}, \citenamefont {Shao},\ and\ \citenamefont
  {Liu}}]{li2024longshortrange}%
  \BibitemOpen
  \bibfield  {author} {\bibinfo {author} {\bibfnamefont {Y.}~\bibnamefont
  {Li}}, \bibinfo {author} {\bibfnamefont {Y.}~\bibnamefont {Wang}}, \bibinfo
  {author} {\bibfnamefont {L.}~\bibnamefont {Huang}}, \bibinfo {author}
  {\bibfnamefont {H.}~\bibnamefont {Yang}}, \bibinfo {author} {\bibfnamefont
  {X.}~\bibnamefont {Wei}}, \bibinfo {author} {\bibfnamefont {J.}~\bibnamefont
  {Zhang}}, \bibinfo {author} {\bibfnamefont {T.}~\bibnamefont {Wang}},
  \bibinfo {author} {\bibfnamefont {Z.}~\bibnamefont {Wang}}, \bibinfo {author}
  {\bibfnamefont {B.}~\bibnamefont {Shao}},\ and\ \bibinfo {author}
  {\bibfnamefont {T.-Y.}\ \bibnamefont {Liu}},\ }in\ \href@noop {} {\emph
  {\bibinfo {booktitle} {The Twelfth International Conference on Learning
  Representations}}}\ (\bibinfo {year} {2024})\BibitemShut {NoStop}%
\bibitem [{\citenamefont {Wu}\ \emph {et~al.}(2024)\citenamefont {Wu},
  \citenamefont {Wu}, \citenamefont {Liu}, \citenamefont {Liu}, \citenamefont
  {Shao},\ and\ \citenamefont {Wang}}]{wu2024se3set}%
  \BibitemOpen
  \bibfield  {author} {\bibinfo {author} {\bibfnamefont {H.}~\bibnamefont
  {Wu}}, \bibinfo {author} {\bibfnamefont {L.}~\bibnamefont {Wu}}, \bibinfo
  {author} {\bibfnamefont {G.}~\bibnamefont {Liu}}, \bibinfo {author}
  {\bibfnamefont {Z.}~\bibnamefont {Liu}}, \bibinfo {author} {\bibfnamefont
  {B.}~\bibnamefont {Shao}},\ and\ \bibinfo {author} {\bibfnamefont
  {Z.}~\bibnamefont {Wang}},\ }\href@noop {} {\bibfield  {journal} {\bibinfo
  {journal} {Preprint at \url{http://arxiv.org/abs/2405.16511}}\ } (\bibinfo
  {year} {2024})}\BibitemShut {NoStop}%
\bibitem [{\citenamefont {Sch{\"u}tt}\ \emph {et~al.}(2019)\citenamefont
  {Sch{\"u}tt}, \citenamefont {Gastegger}, \citenamefont {Tkatchenko},
  \citenamefont {M{\"u}ller},\ and\ \citenamefont
  {Maurer}}]{schutt2019unifying}%
  \BibitemOpen
  \bibfield  {author} {\bibinfo {author} {\bibfnamefont {K.~T.}\ \bibnamefont
  {Sch{\"u}tt}}, \bibinfo {author} {\bibfnamefont {M.}~\bibnamefont
  {Gastegger}}, \bibinfo {author} {\bibfnamefont {A.}~\bibnamefont
  {Tkatchenko}}, \bibinfo {author} {\bibfnamefont {K.-R.}\ \bibnamefont
  {M{\"u}ller}},\ and\ \bibinfo {author} {\bibfnamefont {R.~J.}\ \bibnamefont
  {Maurer}},\ }\href@noop {} {\bibfield  {journal} {\bibinfo  {journal} {Nat.
  Commun.}\ }\textbf {\bibinfo {volume} {10}},\ \bibinfo {pages} {5024}
  (\bibinfo {year} {2019})}\BibitemShut {NoStop}%
\bibitem [{\citenamefont {Unke}\ \emph {et~al.}(2021)\citenamefont {Unke},
  \citenamefont {Bogojeski}, \citenamefont {Gastegger}, \citenamefont {Geiger},
  \citenamefont {Smidt},\ and\ \citenamefont {M{\"u}ller}}]{unke2021se}%
  \BibitemOpen
  \bibfield  {author} {\bibinfo {author} {\bibfnamefont {O.}~\bibnamefont
  {Unke}}, \bibinfo {author} {\bibfnamefont {M.}~\bibnamefont {Bogojeski}},
  \bibinfo {author} {\bibfnamefont {M.}~\bibnamefont {Gastegger}}, \bibinfo
  {author} {\bibfnamefont {M.}~\bibnamefont {Geiger}}, \bibinfo {author}
  {\bibfnamefont {T.}~\bibnamefont {Smidt}},\ and\ \bibinfo {author}
  {\bibfnamefont {K.-R.}\ \bibnamefont {M{\"u}ller}},\ }\href@noop {}
  {\bibfield  {journal} {\bibinfo  {journal} {Advances in Neural Information
  Processing Systems}\ }\textbf {\bibinfo {volume} {34}},\ \bibinfo {pages}
  {14434} (\bibinfo {year} {2021})}\BibitemShut {NoStop}%
\bibitem [{\citenamefont {Li}\ \emph {et~al.}(2022)\citenamefont {Li},
  \citenamefont {Wang}, \citenamefont {Zou}, \citenamefont {Ye}, \citenamefont
  {Xu}, \citenamefont {Gong}, \citenamefont {Duan},\ and\ \citenamefont
  {Xu}}]{li2022deep}%
  \BibitemOpen
  \bibfield  {author} {\bibinfo {author} {\bibfnamefont {H.}~\bibnamefont
  {Li}}, \bibinfo {author} {\bibfnamefont {Z.}~\bibnamefont {Wang}}, \bibinfo
  {author} {\bibfnamefont {N.}~\bibnamefont {Zou}}, \bibinfo {author}
  {\bibfnamefont {M.}~\bibnamefont {Ye}}, \bibinfo {author} {\bibfnamefont
  {R.}~\bibnamefont {Xu}}, \bibinfo {author} {\bibfnamefont {X.}~\bibnamefont
  {Gong}}, \bibinfo {author} {\bibfnamefont {W.}~\bibnamefont {Duan}},\ and\
  \bibinfo {author} {\bibfnamefont {Y.}~\bibnamefont {Xu}},\ }\href@noop {}
  {\bibfield  {journal} {\bibinfo  {journal} {Nat. Comput. Sci.}\ }\textbf
  {\bibinfo {volume} {2}},\ \bibinfo {pages} {367} (\bibinfo {year}
  {2022})}\BibitemShut {NoStop}%
\bibitem [{\citenamefont {Gong}\ \emph {et~al.}(2023)\citenamefont {Gong},
  \citenamefont {Li}, \citenamefont {Zou}, \citenamefont {Xu}, \citenamefont
  {Duan},\ and\ \citenamefont {Xu}}]{gong2023general}%
  \BibitemOpen
  \bibfield  {author} {\bibinfo {author} {\bibfnamefont {X.}~\bibnamefont
  {Gong}}, \bibinfo {author} {\bibfnamefont {H.}~\bibnamefont {Li}}, \bibinfo
  {author} {\bibfnamefont {N.}~\bibnamefont {Zou}}, \bibinfo {author}
  {\bibfnamefont {R.}~\bibnamefont {Xu}}, \bibinfo {author} {\bibfnamefont
  {W.}~\bibnamefont {Duan}},\ and\ \bibinfo {author} {\bibfnamefont
  {Y.}~\bibnamefont {Xu}},\ }\href@noop {} {\bibfield  {journal} {\bibinfo
  {journal} {Nat. Commun.}\ }\textbf {\bibinfo {volume} {14}},\ \bibinfo
  {pages} {2848} (\bibinfo {year} {2023})}\BibitemShut {NoStop}%
\bibitem [{\citenamefont {Yu}\ \emph {et~al.}(2023)\citenamefont {Yu},
  \citenamefont {Xu}, \citenamefont {Qian}, \citenamefont {Qian},\ and\
  \citenamefont {Ji}}]{yu2023efficient}%
  \BibitemOpen
  \bibfield  {author} {\bibinfo {author} {\bibfnamefont {H.}~\bibnamefont
  {Yu}}, \bibinfo {author} {\bibfnamefont {Z.}~\bibnamefont {Xu}}, \bibinfo
  {author} {\bibfnamefont {X.}~\bibnamefont {Qian}}, \bibinfo {author}
  {\bibfnamefont {X.}~\bibnamefont {Qian}},\ and\ \bibinfo {author}
  {\bibfnamefont {S.}~\bibnamefont {Ji}},\ }in\ \href@noop {} {\emph {\bibinfo
  {booktitle} {International Conference on Machine Learning}}}\ (\bibinfo
  {organization} {PMLR},\ \bibinfo {year} {2023})\ pp.\ \bibinfo {pages}
  {40412--40424}\BibitemShut {NoStop}%
\bibitem [{\citenamefont {Zhang}\ \emph {et~al.}()\citenamefont {Zhang},
  \citenamefont {Liu}, \citenamefont {Wang}, \citenamefont {Wei}, \citenamefont
  {Liu}, \citenamefont {Zheng}, \citenamefont {Shao},\ and\ \citenamefont
  {Liu}}]{zhangself}%
  \BibitemOpen
  \bibfield  {author} {\bibinfo {author} {\bibfnamefont {H.}~\bibnamefont
  {Zhang}}, \bibinfo {author} {\bibfnamefont {C.}~\bibnamefont {Liu}}, \bibinfo
  {author} {\bibfnamefont {Z.}~\bibnamefont {Wang}}, \bibinfo {author}
  {\bibfnamefont {X.}~\bibnamefont {Wei}}, \bibinfo {author} {\bibfnamefont
  {S.}~\bibnamefont {Liu}}, \bibinfo {author} {\bibfnamefont {N.}~\bibnamefont
  {Zheng}}, \bibinfo {author} {\bibfnamefont {B.}~\bibnamefont {Shao}},\ and\
  \bibinfo {author} {\bibfnamefont {T.-Y.}\ \bibnamefont {Liu}},\ }in\
  \href@noop {} {\emph {\bibinfo {booktitle} {Forty-first International
  Conference on Machine Learning}}}\BibitemShut {NoStop}%
\bibitem [{\citenamefont {Wang}\ \emph
  {et~al.}(2024{\natexlab{b}})\citenamefont {Wang}, \citenamefont {Liu},
  \citenamefont {Zou}, \citenamefont {Zhang}, \citenamefont {Wei},
  \citenamefont {Huang}, \citenamefont {Wu},\ and\ \citenamefont
  {Shao}}]{wang2024infusing}%
  \BibitemOpen
  \bibfield  {author} {\bibinfo {author} {\bibfnamefont {Z.}~\bibnamefont
  {Wang}}, \bibinfo {author} {\bibfnamefont {C.}~\bibnamefont {Liu}}, \bibinfo
  {author} {\bibfnamefont {N.}~\bibnamefont {Zou}}, \bibinfo {author}
  {\bibfnamefont {H.}~\bibnamefont {Zhang}}, \bibinfo {author} {\bibfnamefont
  {X.}~\bibnamefont {Wei}}, \bibinfo {author} {\bibfnamefont {L.}~\bibnamefont
  {Huang}}, \bibinfo {author} {\bibfnamefont {L.}~\bibnamefont {Wu}},\ and\
  \bibinfo {author} {\bibfnamefont {B.}~\bibnamefont {Shao}},\ }in\ \href@noop
  {} {\emph {\bibinfo {booktitle} {The Thirty-eighth Annual Conference on
  Neural Information Processing Systems}}}\ (\bibinfo {year}
  {2024})\BibitemShut {NoStop}%
\bibitem [{\citenamefont {Li}\ \emph {et~al.}(2024{\natexlab{b}})\citenamefont
  {Li}, \citenamefont {Tang}, \citenamefont {Chen}, \citenamefont {Sun},
  \citenamefont {Zhao}, \citenamefont {Li}, \citenamefont {Tao}, \citenamefont
  {Yuan}, \citenamefont {Duan},\ and\ \citenamefont
  {Xu}}]{li_neural-network_2024}%
  \BibitemOpen
  \bibfield  {author} {\bibinfo {author} {\bibfnamefont {Y.}~\bibnamefont
  {Li}}, \bibinfo {author} {\bibfnamefont {Z.}~\bibnamefont {Tang}}, \bibinfo
  {author} {\bibfnamefont {Z.}~\bibnamefont {Chen}}, \bibinfo {author}
  {\bibfnamefont {M.}~\bibnamefont {Sun}}, \bibinfo {author} {\bibfnamefont
  {B.}~\bibnamefont {Zhao}}, \bibinfo {author} {\bibfnamefont {H.}~\bibnamefont
  {Li}}, \bibinfo {author} {\bibfnamefont {H.}~\bibnamefont {Tao}}, \bibinfo
  {author} {\bibfnamefont {Z.}~\bibnamefont {Yuan}}, \bibinfo {author}
  {\bibfnamefont {W.}~\bibnamefont {Duan}},\ and\ \bibinfo {author}
  {\bibfnamefont {Y.}~\bibnamefont {Xu}},\ }\href
  {https://doi.org/10.1103/PhysRevLett.133.076401} {\bibfield  {journal}
  {\bibinfo  {journal} {Phys. Rev. Lett.}\ }\textbf {\bibinfo {volume} {133}},\
  \bibinfo {pages} {076401} (\bibinfo {year} {2024}{\natexlab{b}})}\BibitemShut
  {NoStop}%
\bibitem [{\citenamefont {Li}\ \emph {et~al.}(2024{\natexlab{c}})\citenamefont
  {Li}, \citenamefont {Tang}, \citenamefont {Fu}, \citenamefont {Dong},
  \citenamefont {Zou}, \citenamefont {Gong}, \citenamefont {Duan},\ and\
  \citenamefont {Xu}}]{li2024deep}%
  \BibitemOpen
  \bibfield  {author} {\bibinfo {author} {\bibfnamefont {H.}~\bibnamefont
  {Li}}, \bibinfo {author} {\bibfnamefont {Z.}~\bibnamefont {Tang}}, \bibinfo
  {author} {\bibfnamefont {J.}~\bibnamefont {Fu}}, \bibinfo {author}
  {\bibfnamefont {W.-H.}\ \bibnamefont {Dong}}, \bibinfo {author}
  {\bibfnamefont {N.}~\bibnamefont {Zou}}, \bibinfo {author} {\bibfnamefont
  {X.}~\bibnamefont {Gong}}, \bibinfo {author} {\bibfnamefont {W.}~\bibnamefont
  {Duan}},\ and\ \bibinfo {author} {\bibfnamefont {Y.}~\bibnamefont {Xu}},\
  }\href@noop {} {\bibfield  {journal} {\bibinfo  {journal} {Phys. Rev. Lett.}\
  }\textbf {\bibinfo {volume} {132}},\ \bibinfo {pages} {096401} (\bibinfo
  {year} {2024}{\natexlab{c}})}\BibitemShut {NoStop}%
\bibitem [{\citenamefont {Zhong}\ \emph {et~al.}(2024)\citenamefont {Zhong},
  \citenamefont {Liu}, \citenamefont {Zhang}, \citenamefont {Tao},
  \citenamefont {Sun}, \citenamefont {Chu}, \citenamefont {Gong}, \citenamefont
  {Yang},\ and\ \citenamefont {Xiang}}]{zhong2024accelerating}%
  \BibitemOpen
  \bibfield  {author} {\bibinfo {author} {\bibfnamefont {Y.}~\bibnamefont
  {Zhong}}, \bibinfo {author} {\bibfnamefont {S.}~\bibnamefont {Liu}}, \bibinfo
  {author} {\bibfnamefont {B.}~\bibnamefont {Zhang}}, \bibinfo {author}
  {\bibfnamefont {Z.}~\bibnamefont {Tao}}, \bibinfo {author} {\bibfnamefont
  {Y.}~\bibnamefont {Sun}}, \bibinfo {author} {\bibfnamefont {W.}~\bibnamefont
  {Chu}}, \bibinfo {author} {\bibfnamefont {X.-G.}\ \bibnamefont {Gong}},
  \bibinfo {author} {\bibfnamefont {J.-H.}\ \bibnamefont {Yang}},\ and\
  \bibinfo {author} {\bibfnamefont {H.}~\bibnamefont {Xiang}},\ }\href@noop {}
  {\bibfield  {journal} {\bibinfo  {journal} {Nat. Comput. Sci.}\ }\textbf
  {\bibinfo {volume} {4}},\ \bibinfo {pages} {615} (\bibinfo {year}
  {2024})}\BibitemShut {NoStop}%
\bibitem [{\citenamefont {Karsai}\ \emph {et~al.}(2018)\citenamefont {Karsai},
  \citenamefont {Engel}, \citenamefont {Flage-Larsen},\ and\ \citenamefont
  {Kresse}}]{karsai2018electron}%
  \BibitemOpen
  \bibfield  {author} {\bibinfo {author} {\bibfnamefont {F.}~\bibnamefont
  {Karsai}}, \bibinfo {author} {\bibfnamefont {M.}~\bibnamefont {Engel}},
  \bibinfo {author} {\bibfnamefont {E.}~\bibnamefont {Flage-Larsen}},\ and\
  \bibinfo {author} {\bibfnamefont {G.}~\bibnamefont {Kresse}},\ }\href@noop {}
  {\bibfield  {journal} {\bibinfo  {journal} {New J. Phys.}\ }\textbf {\bibinfo
  {volume} {20}},\ \bibinfo {pages} {123008} (\bibinfo {year}
  {2018})}\BibitemShut {NoStop}%
\bibitem [{\citenamefont {Gong}\ \emph {et~al.}(2024)\citenamefont {Gong},
  \citenamefont {Louie}, \citenamefont {Duan},\ and\ \citenamefont
  {Xu}}]{gong2024generalizing}%
  \BibitemOpen
  \bibfield  {author} {\bibinfo {author} {\bibfnamefont {X.}~\bibnamefont
  {Gong}}, \bibinfo {author} {\bibfnamefont {S.~G.}\ \bibnamefont {Louie}},
  \bibinfo {author} {\bibfnamefont {W.}~\bibnamefont {Duan}},\ and\ \bibinfo
  {author} {\bibfnamefont {Y.}~\bibnamefont {Xu}},\ }\href@noop {} {\bibfield
  {journal} {\bibinfo  {journal} {Nat. Comput. Sci.}\ }\textbf {\bibinfo
  {volume} {4}},\ \bibinfo {pages} {752} (\bibinfo {year} {2024})}\BibitemShut
  {NoStop}%
\bibitem [{\citenamefont {Tang}\ \emph {et~al.}(2024)\citenamefont {Tang},
  \citenamefont {Li}, \citenamefont {Lin}, \citenamefont {Gong}, \citenamefont
  {Jin}, \citenamefont {He}, \citenamefont {Jiang}, \citenamefont {Ren},
  \citenamefont {Duan},\ and\ \citenamefont {Xu}}]{tang2024deep}%
  \BibitemOpen
  \bibfield  {author} {\bibinfo {author} {\bibfnamefont {Z.}~\bibnamefont
  {Tang}}, \bibinfo {author} {\bibfnamefont {H.}~\bibnamefont {Li}}, \bibinfo
  {author} {\bibfnamefont {P.}~\bibnamefont {Lin}}, \bibinfo {author}
  {\bibfnamefont {X.}~\bibnamefont {Gong}}, \bibinfo {author} {\bibfnamefont
  {G.}~\bibnamefont {Jin}}, \bibinfo {author} {\bibfnamefont {L.}~\bibnamefont
  {He}}, \bibinfo {author} {\bibfnamefont {H.}~\bibnamefont {Jiang}}, \bibinfo
  {author} {\bibfnamefont {X.}~\bibnamefont {Ren}}, \bibinfo {author}
  {\bibfnamefont {W.}~\bibnamefont {Duan}},\ and\ \bibinfo {author}
  {\bibfnamefont {Y.}~\bibnamefont {Xu}},\ }\href@noop {} {\bibfield  {journal}
  {\bibinfo  {journal} {Nat. Commun.}\ }\textbf {\bibinfo {volume} {15}},\
  \bibinfo {pages} {8815} (\bibinfo {year} {2024})}\BibitemShut {NoStop}%
\bibitem [{\citenamefont {Bhimanapati}\ \emph {et~al.}(2015)\citenamefont
  {Bhimanapati}, \citenamefont {Lin}, \citenamefont {Meunier}, \citenamefont
  {Jung}, \citenamefont {Cha}, \citenamefont {Das}, \citenamefont {Xiao},
  \citenamefont {Son}, \citenamefont {Strano}, \citenamefont {Cooper} \emph
  {et~al.}}]{bhimanapati2015recent}%
  \BibitemOpen
  \bibfield  {author} {\bibinfo {author} {\bibfnamefont {G.~R.}\ \bibnamefont
  {Bhimanapati}}, \bibinfo {author} {\bibfnamefont {Z.}~\bibnamefont {Lin}},
  \bibinfo {author} {\bibfnamefont {V.}~\bibnamefont {Meunier}}, \bibinfo
  {author} {\bibfnamefont {Y.}~\bibnamefont {Jung}}, \bibinfo {author}
  {\bibfnamefont {J.}~\bibnamefont {Cha}}, \bibinfo {author} {\bibfnamefont
  {S.}~\bibnamefont {Das}}, \bibinfo {author} {\bibfnamefont {D.}~\bibnamefont
  {Xiao}}, \bibinfo {author} {\bibfnamefont {Y.}~\bibnamefont {Son}}, \bibinfo
  {author} {\bibfnamefont {M.~S.}\ \bibnamefont {Strano}}, \bibinfo {author}
  {\bibfnamefont {V.~R.}\ \bibnamefont {Cooper}}, \emph {et~al.},\ }\href@noop
  {} {\bibfield  {journal} {\bibinfo  {journal} {ACS nano}\ }\textbf {\bibinfo
  {volume} {9}},\ \bibinfo {pages} {11509} (\bibinfo {year}
  {2015})}\BibitemShut {NoStop}%
\bibitem [{\citenamefont {Gilmer}\ \emph {et~al.}(2017)\citenamefont {Gilmer},
  \citenamefont {Schoenholz}, \citenamefont {Riley}, \citenamefont {Vinyals},\
  and\ \citenamefont {Dahl}}]{gilmer2017neural}%
  \BibitemOpen
  \bibfield  {author} {\bibinfo {author} {\bibfnamefont {J.}~\bibnamefont
  {Gilmer}}, \bibinfo {author} {\bibfnamefont {S.~S.}\ \bibnamefont
  {Schoenholz}}, \bibinfo {author} {\bibfnamefont {P.~F.}\ \bibnamefont
  {Riley}}, \bibinfo {author} {\bibfnamefont {O.}~\bibnamefont {Vinyals}},\
  and\ \bibinfo {author} {\bibfnamefont {G.~E.}\ \bibnamefont {Dahl}},\ }in\
  \href@noop {} {\emph {\bibinfo {booktitle} {International conference on
  machine learning}}}\ (\bibinfo {organization} {PMLR},\ \bibinfo {year}
  {2017})\ pp.\ \bibinfo {pages} {1263--1272}\BibitemShut {NoStop}%
\bibitem [{\citenamefont {Lucignano}\ \emph {et~al.}(2019)\citenamefont
  {Lucignano}, \citenamefont {Alf{\`e}}, \citenamefont {Cataudella},
  \citenamefont {Ninno},\ and\ \citenamefont {Cantele}}]{lucignano2019crucial}%
  \BibitemOpen
  \bibfield  {author} {\bibinfo {author} {\bibfnamefont {P.}~\bibnamefont
  {Lucignano}}, \bibinfo {author} {\bibfnamefont {D.}~\bibnamefont {Alf{\`e}}},
  \bibinfo {author} {\bibfnamefont {V.}~\bibnamefont {Cataudella}}, \bibinfo
  {author} {\bibfnamefont {D.}~\bibnamefont {Ninno}},\ and\ \bibinfo {author}
  {\bibfnamefont {G.}~\bibnamefont {Cantele}},\ }\href@noop {} {\bibfield
  {journal} {\bibinfo  {journal} {Phys. Rev. B}\ }\textbf {\bibinfo {volume}
  {99}},\ \bibinfo {pages} {195419} (\bibinfo {year} {2019})}\BibitemShut
  {NoStop}%
\bibitem [{\citenamefont {Kresse}\ and\ \citenamefont
  {Hafner}(1993)}]{kresse1993ab}%
  \BibitemOpen
  \bibfield  {author} {\bibinfo {author} {\bibfnamefont {G.}~\bibnamefont
  {Kresse}}\ and\ \bibinfo {author} {\bibfnamefont {J.}~\bibnamefont
  {Hafner}},\ }\href@noop {} {\bibfield  {journal} {\bibinfo  {journal} {Phys.
  Rev. B}\ }\textbf {\bibinfo {volume} {47}},\ \bibinfo {pages} {558} (\bibinfo
  {year} {1993})}\BibitemShut {NoStop}%
\bibitem [{\citenamefont {Kresse}\ and\ \citenamefont
  {Furthm{\"u}ller}(1996)}]{kresse1996computational}%
  \BibitemOpen
  \bibfield  {author} {\bibinfo {author} {\bibfnamefont {G.}~\bibnamefont
  {Kresse}}\ and\ \bibinfo {author} {\bibfnamefont {J.}~\bibnamefont
  {Furthm{\"u}ller}},\ }\href@noop {} {\bibfield  {journal} {\bibinfo
  {journal} {Comput. Mater. Sci.}\ }\textbf {\bibinfo {volume} {6}},\ \bibinfo
  {pages} {15} (\bibinfo {year} {1996})}\BibitemShut {NoStop}%
\bibitem [{\citenamefont {Kresse}\ and\ \citenamefont
  {Joubert}(1999)}]{kresse1999ultrasoft}%
  \BibitemOpen
  \bibfield  {author} {\bibinfo {author} {\bibfnamefont {G.}~\bibnamefont
  {Kresse}}\ and\ \bibinfo {author} {\bibfnamefont {D.}~\bibnamefont
  {Joubert}},\ }\href@noop {} {\bibfield  {journal} {\bibinfo  {journal} {Phys.
  Rev. B}\ }\textbf {\bibinfo {volume} {59}},\ \bibinfo {pages} {1758}
  (\bibinfo {year} {1999})}\BibitemShut {NoStop}%
\bibitem [{\citenamefont {Perdew}\ \emph {et~al.}(1996)\citenamefont {Perdew},
  \citenamefont {Burke},\ and\ \citenamefont
  {Ernzerhof}}]{perdew1996generalized}%
  \BibitemOpen
  \bibfield  {author} {\bibinfo {author} {\bibfnamefont {J.~P.}\ \bibnamefont
  {Perdew}}, \bibinfo {author} {\bibfnamefont {K.}~\bibnamefont {Burke}},\ and\
  \bibinfo {author} {\bibfnamefont {M.}~\bibnamefont {Ernzerhof}},\ }\href@noop
  {} {\bibfield  {journal} {\bibinfo  {journal} {Phys. Rev. Lett.}\ }\textbf
  {\bibinfo {volume} {77}},\ \bibinfo {pages} {3865} (\bibinfo {year}
  {1996})}\BibitemShut {NoStop}%
\bibitem [{\citenamefont {Grimme}(2006)}]{grimme2006semiempirical}%
  \BibitemOpen
  \bibfield  {author} {\bibinfo {author} {\bibfnamefont {S.}~\bibnamefont
  {Grimme}},\ }\href@noop {} {\bibfield  {journal} {\bibinfo  {journal} {J.
  Comput. Chem.}\ }\textbf {\bibinfo {volume} {27}},\ \bibinfo {pages} {1787}
  (\bibinfo {year} {2006})}\BibitemShut {NoStop}%
\bibitem [{\citenamefont {Ozaki}(2003)}]{ozaki2003variationally}%
  \BibitemOpen
  \bibfield  {author} {\bibinfo {author} {\bibfnamefont {T.}~\bibnamefont
  {Ozaki}},\ }\href@noop {} {\bibfield  {journal} {\bibinfo  {journal} {Phys.
  Rev. B}\ }\textbf {\bibinfo {volume} {67}},\ \bibinfo {pages} {155108}
  (\bibinfo {year} {2003})}\BibitemShut {NoStop}%
\bibitem [{\citenamefont {Ozaki}\ and\ \citenamefont
  {Kino}(2004)}]{ozaki2004numerical}%
  \BibitemOpen
  \bibfield  {author} {\bibinfo {author} {\bibfnamefont {T.}~\bibnamefont
  {Ozaki}}\ and\ \bibinfo {author} {\bibfnamefont {H.}~\bibnamefont {Kino}},\
  }\href@noop {} {\bibfield  {journal} {\bibinfo  {journal} {Phys. Rev. B}\
  }\textbf {\bibinfo {volume} {69}},\ \bibinfo {pages} {195113} (\bibinfo
  {year} {2004})}\BibitemShut {NoStop}%
\bibitem [{\citenamefont {Attaccalite}\ \emph {et~al.}(2010)\citenamefont
  {Attaccalite}, \citenamefont {Wirtz}, \citenamefont {Lazzeri}, \citenamefont
  {Mauri},\ and\ \citenamefont {Rubio}}]{attaccalite_doped_2010}%
  \BibitemOpen
  \bibfield  {author} {\bibinfo {author} {\bibfnamefont {C.}~\bibnamefont
  {Attaccalite}}, \bibinfo {author} {\bibfnamefont {L.}~\bibnamefont {Wirtz}},
  \bibinfo {author} {\bibfnamefont {M.}~\bibnamefont {Lazzeri}}, \bibinfo
  {author} {\bibfnamefont {F.}~\bibnamefont {Mauri}},\ and\ \bibinfo {author}
  {\bibfnamefont {A.}~\bibnamefont {Rubio}},\ }\href
  {https://doi.org/10.1021/nl9034626} {\bibfield  {journal} {\bibinfo
  {journal} {Nano Lett.}\ }\textbf {\bibinfo {volume} {10}},\ \bibinfo {pages}
  {1172} (\bibinfo {year} {2010})}\BibitemShut {NoStop}%
\bibitem [{\citenamefont {Sohier}\ \emph {et~al.}(2014)\citenamefont {Sohier},
  \citenamefont {Calandra}, \citenamefont {Park}, \citenamefont {Bonini},
  \citenamefont {Marzari},\ and\ \citenamefont
  {Mauri}}]{sohier_phonon-limited_2014}%
  \BibitemOpen
  \bibfield  {author} {\bibinfo {author} {\bibfnamefont {T.}~\bibnamefont
  {Sohier}}, \bibinfo {author} {\bibfnamefont {M.}~\bibnamefont {Calandra}},
  \bibinfo {author} {\bibfnamefont {C.-H.}\ \bibnamefont {Park}}, \bibinfo
  {author} {\bibfnamefont {N.}~\bibnamefont {Bonini}}, \bibinfo {author}
  {\bibfnamefont {N.}~\bibnamefont {Marzari}},\ and\ \bibinfo {author}
  {\bibfnamefont {F.}~\bibnamefont {Mauri}},\ }\href
  {https://doi.org/10.1103/PhysRevB.90.125414} {\bibfield  {journal} {\bibinfo
  {journal} {Phys. Rev. B}\ }\textbf {\bibinfo {volume} {90}},\ \bibinfo
  {pages} {125414} (\bibinfo {year} {2014})}\BibitemShut {NoStop}%
\bibitem [{exa()}]{example}%
  \BibitemOpen
  \href@noop {} {}\bibinfo {note} {For undoped monolayer graphene, the Fermi
  velocity is predicted to be reduced by approximately 20\%; applying the same
  method to bilayer graphene should yield a comparable estimate
  \cite{park_electronphonon_2008}.}\BibitemShut {Stop}%
\bibitem [{\citenamefont {Wang}\ \emph
  {et~al.}(2022{\natexlab{b}})\citenamefont {Wang}, \citenamefont {Wang},
  \citenamefont {Zhao}, \citenamefont {Xu}, \citenamefont {Hao}, \citenamefont
  {Hsieh}, \citenamefont {Gu},\ and\ \citenamefont
  {Duan}}]{wang2022heterogeneous}%
  \BibitemOpen
  \bibfield  {author} {\bibinfo {author} {\bibfnamefont {Z.}~\bibnamefont
  {Wang}}, \bibinfo {author} {\bibfnamefont {C.}~\bibnamefont {Wang}}, \bibinfo
  {author} {\bibfnamefont {S.}~\bibnamefont {Zhao}}, \bibinfo {author}
  {\bibfnamefont {Y.}~\bibnamefont {Xu}}, \bibinfo {author} {\bibfnamefont
  {S.}~\bibnamefont {Hao}}, \bibinfo {author} {\bibfnamefont {C.~Y.}\
  \bibnamefont {Hsieh}}, \bibinfo {author} {\bibfnamefont {B.-L.}\ \bibnamefont
  {Gu}},\ and\ \bibinfo {author} {\bibfnamefont {W.}~\bibnamefont {Duan}},\
  }\href@noop {} {\bibfield  {journal} {\bibinfo  {journal} {npj Comput.
  Mater.}\ }\textbf {\bibinfo {volume} {8}},\ \bibinfo {pages} {53} (\bibinfo
  {year} {2022}{\natexlab{b}})}\BibitemShut {NoStop}%
\bibitem [{\citenamefont {Park}\ \emph {et~al.}(2008)\citenamefont {Park},
  \citenamefont {Giustino}, \citenamefont {Cohen},\ and\ \citenamefont
  {Louie}}]{park_electronphonon_2008}%
  \BibitemOpen
  \bibfield  {author} {\bibinfo {author} {\bibfnamefont {C.-H.}\ \bibnamefont
  {Park}}, \bibinfo {author} {\bibfnamefont {F.}~\bibnamefont {Giustino}},
  \bibinfo {author} {\bibfnamefont {M.~L.}\ \bibnamefont {Cohen}},\ and\
  \bibinfo {author} {\bibfnamefont {S.~G.}\ \bibnamefont {Louie}},\ }\href
  {https://doi.org/10.1021/nl801884n} {\bibfield  {journal} {\bibinfo
  {journal} {Nano Lett.}\ }\textbf {\bibinfo {volume} {8}},\ \bibinfo {pages}
  {4229} (\bibinfo {year} {2008})}\BibitemShut {NoStop}%
\end{thebibliography}
%

\end{document}